# Real-Time Image Analysis Software

# Suitable for Resource-Constrained Computing

Alexandre Matov[1, †]

[1] DataSet Analysis LLC, 155 Jackson St, San Francisco, CA 94111, United States

[†] Corresponding author:

email: matov@datasetanalysis.com

## ABSTRACT

**Introduction:** Methods for personalizing medical treatment are the focal point of contemporary clinical research. In cancer care, for instance, we can analyze the effects of therapies at the level of individual cells. Complete characterization of treatment efficacy and evaluation of why some individuals respond to specific regimens, whereas others do not, requires additional approaches to genetic sequencing at single time points. Methods for the continuous analysis of changes in phenotype, such as morphology and motion tracking of cellular proteins and organelles over time frames spanning the minute-hour scales, can provide important insight to patient treatment options. The integration of measurements of intracellular dynamics and the contribution of multiple genetic pathways in degenerative diseases is vital for the development of biomarkers for the early detection of pathogenesis and therapy efficacy.

**Methods:** We have developed a software suite (DataSet Tracker) for real-time analysis designed to run on computers, smartphones, and smart glasses hardware and suitable for resource-constrained, on-the-fly computing in microscopes without internet connectivity; a demo is available for viewing in Vid. 1. Our objective is to present the community with an integrated, easy to use by all, tool for resolving the complex dynamics of the cytoskeletal meshworks, intracytoplasmic membranous networks, and vesicle trafficking. Our software is optimized for resource-constrained computing and can be installed even on microscopes without internet connectivity.

**Results:** Our computational platform can provide high-content analyses and functional secondary screening of novel compounds that are in the process of approval, or at a pre-clinical stage of development, and putative combination therapies based on FDA-approved drugs. Importantly, dissecting the mechanisms of drug action with quantitative detail will allow the design of drugs that impede relapse and optimal dose regimens with minimal harmful side effects by carefully exploiting disease-specific aberrations.

**Conclusions:** DataSet Tracker, the real-time optical flow feature tracking software presented in this contribution, can serve as the base module of an integrated platform of existing and future algorithms for real-time cellular analysis. The computational assay we propose could successfully be applied to evaluate treatment strategies for any human organ. It is our goal to have this integrated tool approved for use in the clinical practice.

## INTRODUCTION

We present software for real-time image analysis. The algorithms were tested on personal computers, cell phones, and smart glasses. The objective of this work is to facilitate research work, primarily in live-cell biology, during which scientists are required to utilize multi-step customized processes in order to analyze imaging data. All existing software packages analyze images off-line, meaning that researchers collect imaging datasets, for example using high-resolution microscopy, and then transfer the data, often comprising large size files, to another computer where they need access to specialized software tools for different analyses.

Our interest is to streamline this process in two ways. (1) By offering real-time analysis tools, which - because there will no longer be a need to transfer data - would save an enormous amount of time. Our objective is to integrate different computer vision modules in one single package with multiple analysis modules that perform a variety of functions. Such integration will eliminate the need to load imaging data in different software tools one by one sequentially, as it is often done in science labs at present. (2) Because our software displays analysis results in real time (5 frames per second for the examples shown in this manuscript), it would allow us to improve the calibration of experimental conditions, such as drug treatment regimens. The reason for this is that the changes in the behavior of living cells can be measured, with quantitative detail, during sample manipulation and image acquisition. These measurements will allow us for the first time to identify subtle changes in biological function, which cannot be observed by the naked eye. This way, the real-time software will enhance our understanding of human physiology and the regulation of individual cells.

The complex dynamics of cytoskeletal proteins make them a difficult subject of study without quantitative analytical tools. The *ex vivo* analysis of living patient-derived cells comprises of the evaluation of the response of cytoskeletal filaments and meshworks, intracytoplasmic membranous networks, and vesicle trafficking before and after drug treatment, and requires motion tracking. Tracking is the process of finding object (feature) motion correspondence from one time-lapse frame to the next throughout a time-lapse sequence. The features can be fluorescently labeled proteins or imaged label-free using a microscopy technique that highlights the gray-scale intensity gradient to their immediate surroundings in the image. Methods that can reliably analyze the evolution of the morphology and localization of cellular proteins over hundreds of time-lapse frames will be very relevant for capturing the changes in the subcellular organization in disease and during treatment (Matov, 2024d; Matov, 2024f; Matov, 2025a; Matov, 2025b; Matov et al., 2010; Matov et al., 2011). Such methods can allow physicians to compare visually and quantitatively the effects of treatment regimens and select the one most likely to be efficacious.

Furthermore, intercellular dynamic behavior can be analyzed with available computer vision robotics libraries, which deliver results in real-time. On-the-fly image analysis and instant quantitative feedback can significantly speed up clinical work, and allow for the precise calibration of the imaging set-up and optimization of the drug regimens based on considerations of their effects on cell function. To improve on the ability to precisely calibrate the pharmacological set-up, augmented reality artificial intelligence (AI) software can be added to existing live-cell microscopes, thus providing instant feedback during sample observation and image acquisition. The selection of drug dose and deducing its effects (Milas et al., 1995; Skipper, 1971; Zasadil et al., 2014), depending on the dose administered, is a daunting task.

During treatment of tubulin inhibitors, for instance, different microtubule (MT)-associated proteins (Vale, 1987) are activated depending on the drug dose administered (Dhamodharan et al., 1995; Jordan et al., 1992; Jordan et al., 1993; Jordan and Wilson, 1990; Toso et al., 1993), and consequently, different resistance mechanisms (Torres and Horwitz, 1998) may be triggered. The effects of MT inhibitors change nonlinearly with a dose increase (Potashnikova et al., 2019). While a high dose of nocodazole depolymerizes MTs, a low dose, which could lead to no side effects, increases polymerization rates (Thoma et al., 2010). Also, while a high dose of paclitaxel stabilizes MTs, a low dose similarly increases MT polymerization rates (Matov, 2024f). To better elucidate the mechanisms behind the non-linearity of drug action, real-time analysis software will allow us to store tracking results during sample observation and, this way, it will allow the detailed documentation and quantification of all observations made by a scientist.

Our work has been focused on the automated analysis of cytoskeletal dynamics, predominantly those of MTs (Gatlin et al., 2010; Gatlin et al., 2009; Houghtaling et al., 2009) and filamentous actin (F-actin) (Adams et al., 2004; Ponti et al., 2005), to study the effects of tubulin, actin, and tropomyosin inhibitors, and post-translational modifications as well as the modulation of focal adhesions (FAs) (Spanjaard et al., 2015) and ectopic activation of GSK3β (Kumar et al., 2009). During pathogenesis, or as a result of drug treatment in disease, cells change their intracellular organization, rearrange their internal components as they grow, divide, and adapt mechanically to a hostile environment. These functions depend on protein filaments (the cytoskeleton and FAs) that provide the cell shape and its capacity for directed movement.

The main novelty presented in the manuscript is related to the ability to quantify and visualize, in real time, the changes in morphology and motion of cellular components as the cell adapts to a wide range of biochemical and mechanical perturbations. We have investigated these changes in the context of the two main cytoskeletal proteins, tubulin and actin. Tubulin has been the target of numerous therapeutic approaches for six decades now. Most of these therapies have been, in essence, a "black box" approach (Bunge, 1963), where a treatment with debilitating side effects is administered without having the ability to anticipate whether it would affect the desired target.

With our approach, an instant quantitative evaluation of some of the more difficult to visually interpret effects of tubulin-targeting compounds can be performed while calibrating the imaging and pharmacological set-up, and during image acquisition, which will facilitate deducing mechanistic explanations for the reasons behind the measured differences. Actin, for instance, is instrumental in wound healing as well as during tumor metastasis; it is also the most important protein in heart and other muscle function. Very few drugs currently modulate actin activity directly, albeit the abundant availability of anti-inflammatory medications that affect the COX pathway (Loerke et al., 2012). It is the variability of patient response in genomics precision medicine that underscores the importance of utilizing quantitative imaging techniques in the clinic.

There is a lot we do not know about the regulation of our organs, the cells comprising them, and the interdependencies between them. The validation of novel prognostic and predictive biomarkers can be done by comparing the *ex vivo* tumor growth kinetics and treatment predictions to outcomes of patients with similar, to primary cells after treatment, genetic profiles, who are enrolled in clinical trials.

## RESULTS

### Segmentation of cellular areas

Taxane treatment induces a variety of changes in cancer cells, including changes in MT polymer density (Foisner and Wiche, 1985; Sung and Giannakakou, 2014). Differences in baseline polymer density exist also between drug-sensitive and drug-resistant cells. To quantify such differences, we quantify the overall MT intensity in images of A549 lung cancer cells before and after knockdown of BRCA1 (B1KD). Because such cells grow in clusters, we used nuclear labeling to segment the areas of the individual cells (see Materials and Methods). The B1KD cells were systematically dimmer and appeared to have a lower level of MT density in comparison to the drug-resistant parental A549 cells (Fig. 1). Our results suggest that reduced MT polymer density correlates with improved sensitivity to taxane treatment in lung cancer.

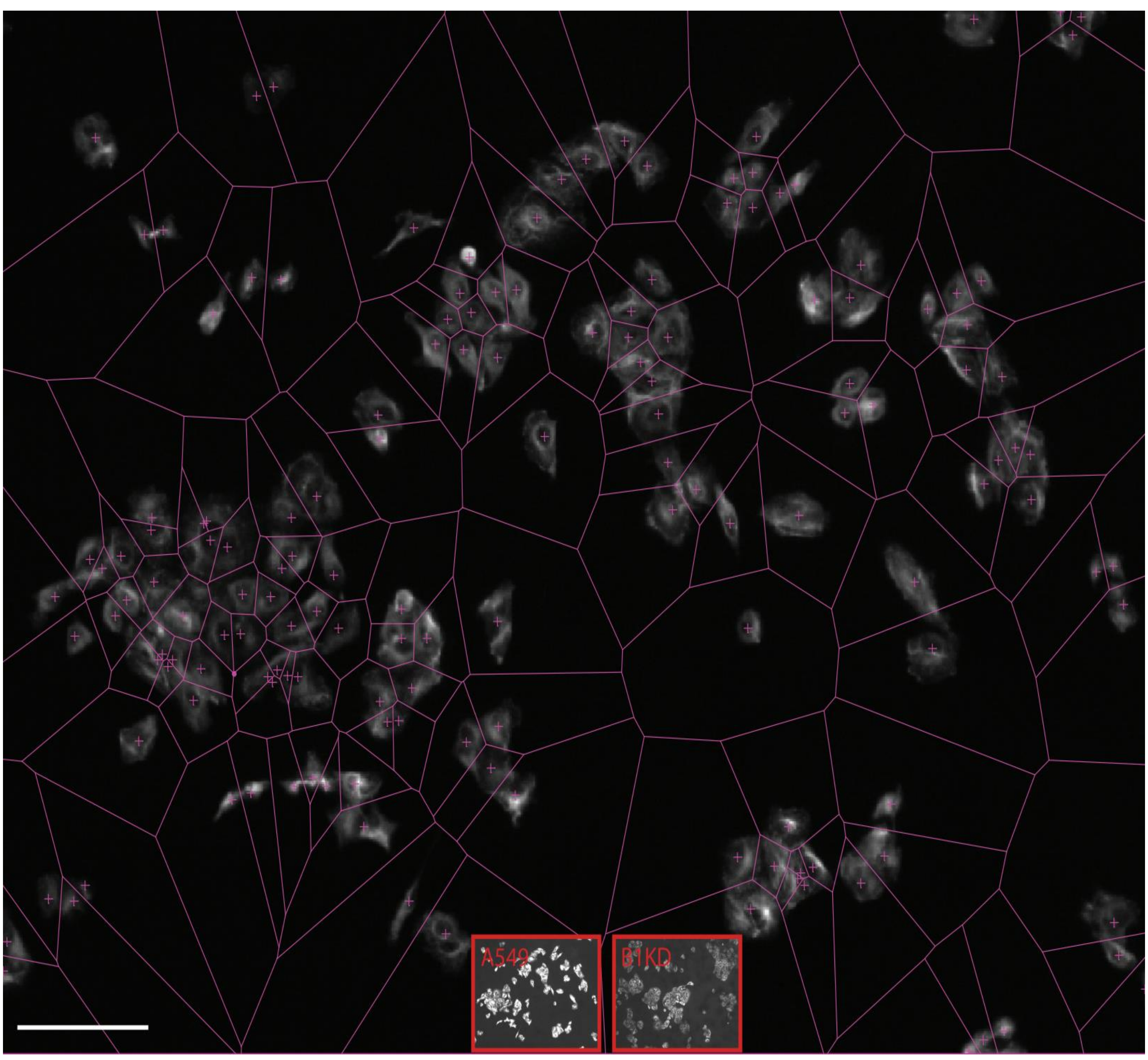

**Figure 1. Segmentation of the area of A549 lung cancer cells.** Tubulin is fluorescently labeled. The figure presents an image segmentation approach to determine the boundaries between cells growing in clusters. Scale bar equals 40 µm. The insets show a comparison between the intensity in parental A549 cells (left inset) and BRCA1 knockdown cells (right inset), which appear dimmer.

To further investigate resistance to treatment with tubulin inhibitors, we performed a sulforhodamine B colorimetric assay for cytotoxicity screening (Vichai and Kirtikara, 2006) in MDA-MB-231 and SK-BR-3 breast cancer cell lines. Low doses of paclitaxel (about 2 nM) and docetaxel (0.2 nM) – contraintuitively - increased with up to 18% (paclitaxel) or 24% (docetaxel) the cell density in these cell lines, indicating increased cell proliferation to result from drug treatment at low concentrations. In addition, the portion of the cytoplasmic area of breast cancer cells was considerably larger compared to the nucleus than in prostate cancer cells (LNCaP) and prostate cancer patients' circulating tumor cells (CTCs), based on prostate-specific membrane antigen (PSMA) and DAPI staining (Fig. 2). This is an indication that the extent of the MT networks varies between different types of epithelial tumors, which indicates susceptibility of different types of tubulin inhibitors, such as polymer destabilizers (like the vinca alkaloids) in lung and breast cancer, and polymer stabilizers (like the taxanes) in androgen-sensitive prostate cancer.

However, our hypothesis is that MT-destabilizing drugs can be very efficient in the treatment of particular prostate tumors – for example in androgen-independent disease (Matov, 2024c) or having other vulnerabilities (Matov, 2024g). In colorectal cancer, vinca alkaloids can be successfully used (Mertens et al., 2023) off-label and BRAF-null tumors are susceptible even in very advanced disease (Vecchione et al., 2016), leading to a complete and lasting remission (Linn et al., 1994). In breast cancer, our research on MT dynamics has indicated that receptor-negative tumors, for which there are no efficient treatment options (Maqbool et al., 2022), exhibit tumor-specific MT dynamics (Matov, 2024e) and might be tentatively susceptible to vinca alkaloids as a first-line of chemotherapy (Matov, 2024f).

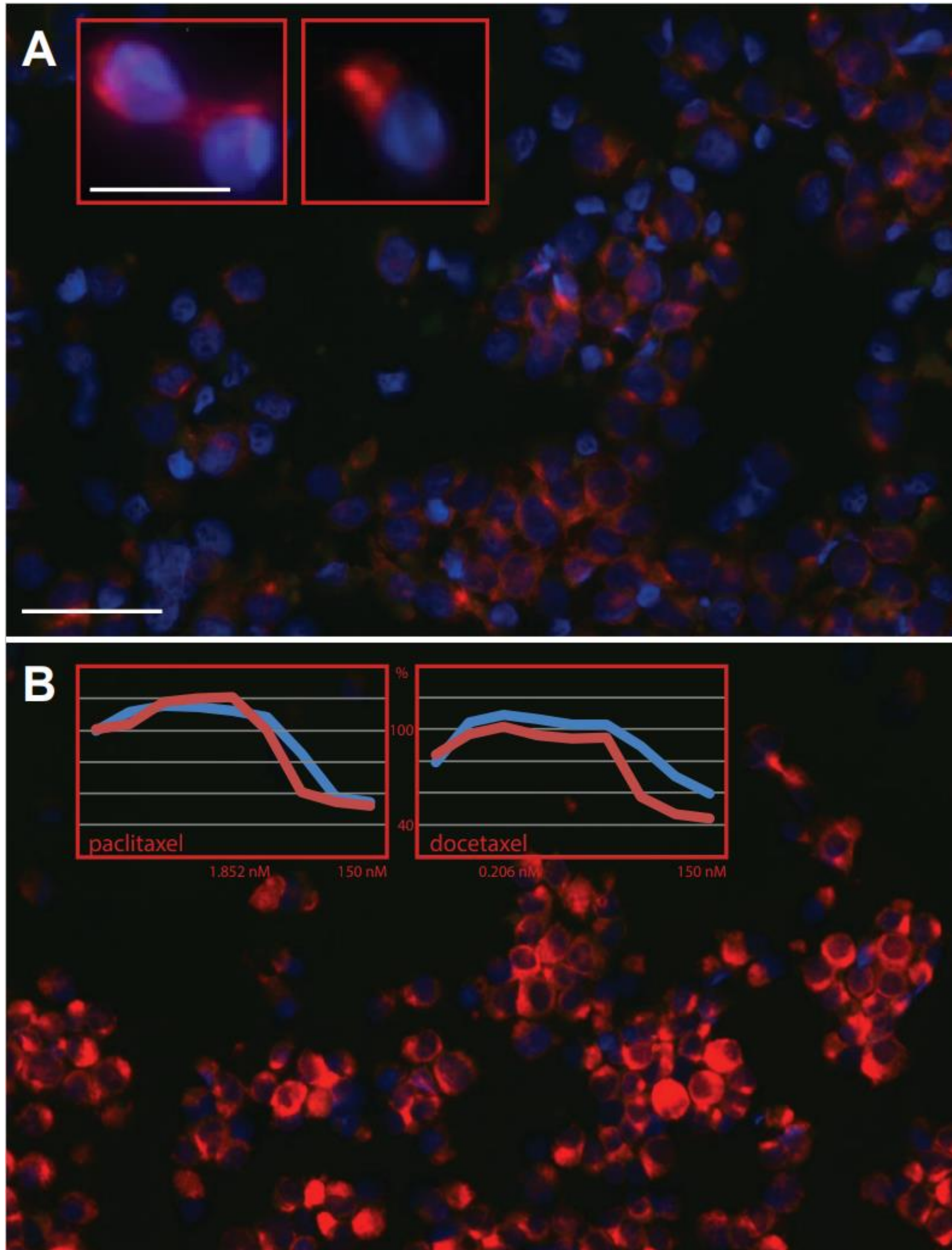

**Figure 2. Cell density and extent of the cytoplasm in cancer cells.** Figure legend: PSMA (J591, tumor marker, red), CD45 (leukocyte marker, green), DAPI (nuclear marker, blue). Scale bar equals 70 µm. (A) LNCaP prostate cancer cells. Inset: Androgen-sensitive prostate cancer CTCs. Inset scale bar equals 20 µm. (B) MDA-MB-231 breast cancer cells. The insets show cell density determination values based on a sulforhodamine B cytotoxicity assay for MDA-MB-231 (blue line) and SK-BR-3 breast cancer (red line) cells after titration with two tubulin inhibitors.

### Microtubules modulate the composition of tumor secretome

Cancer cells activate their surrounding fibroblasts to promote remodeling of their microenvironment (Potenta et al., 2008). Taxane treatment attenuates fibroblast motility in HMF3S cells and reduces the levels of lysyl oxidase homologue 2, c-Met, fibronectin, TGFβ, CTGF, importin α, and 14-3-3 in MDA-MB-231 cells (Tran et al., 2013; Tran et al., 2012). Lowering the levels of fibronectin (Duband et al., 1988) reduces cell spreading as well as lamellipodial protrusion and retraction (Maheshwari et al., 1999), which indicates decreased cellular contractility. Lysyl oxidase-like 2 promotes angiogenesis through modulation of AKT and FA kinase signaling (de Jong et al., 2019). A mutation in RAF that modulates the interaction with the scaffold protein 14-3-3 has been shown to lead to clinical resistance to treatment with the RAS(ON) tri-complex inhibitor (TCI) daraxonrasib (RMC-6236) (Sang et al., 2025).

TGFβ (Blobe et al., 2000; Massagué and Sheppard, 2023) is a key driver of tumor progression and extracellular matrix remodeling. We used lipopolysaccharide (LPS) stimulation to stimulate the secretion of TGFβ (Xie et al., 2009) and analyzed cell membrane-associated TGFβ in populations of both untreated and treated with paclitaxel non-permeabilized cells (see Materials and Methods). Our measurements indicated that in untreated MDA-MB-231 cells, LPS stimulation resulted in an increase of bright foci of TGFβ at the cell membrane, reflecting accumulation of TGFβ-containing secretory granules on the cell surface. In contrast, paclitaxel treated MDA-MB-231 cells did not exhibit increased TGFβ plasma membrane localization upon LPS stimulation. Overall, paclitaxel-mediated stabilization of MTs in breast cancer cells inhibited the trafficking and secretion of TGFβ (Fig. 3). These data indicate that tubulin inhibitors directly affect trafficking of TGFβ-containing granules to the cell membrane.

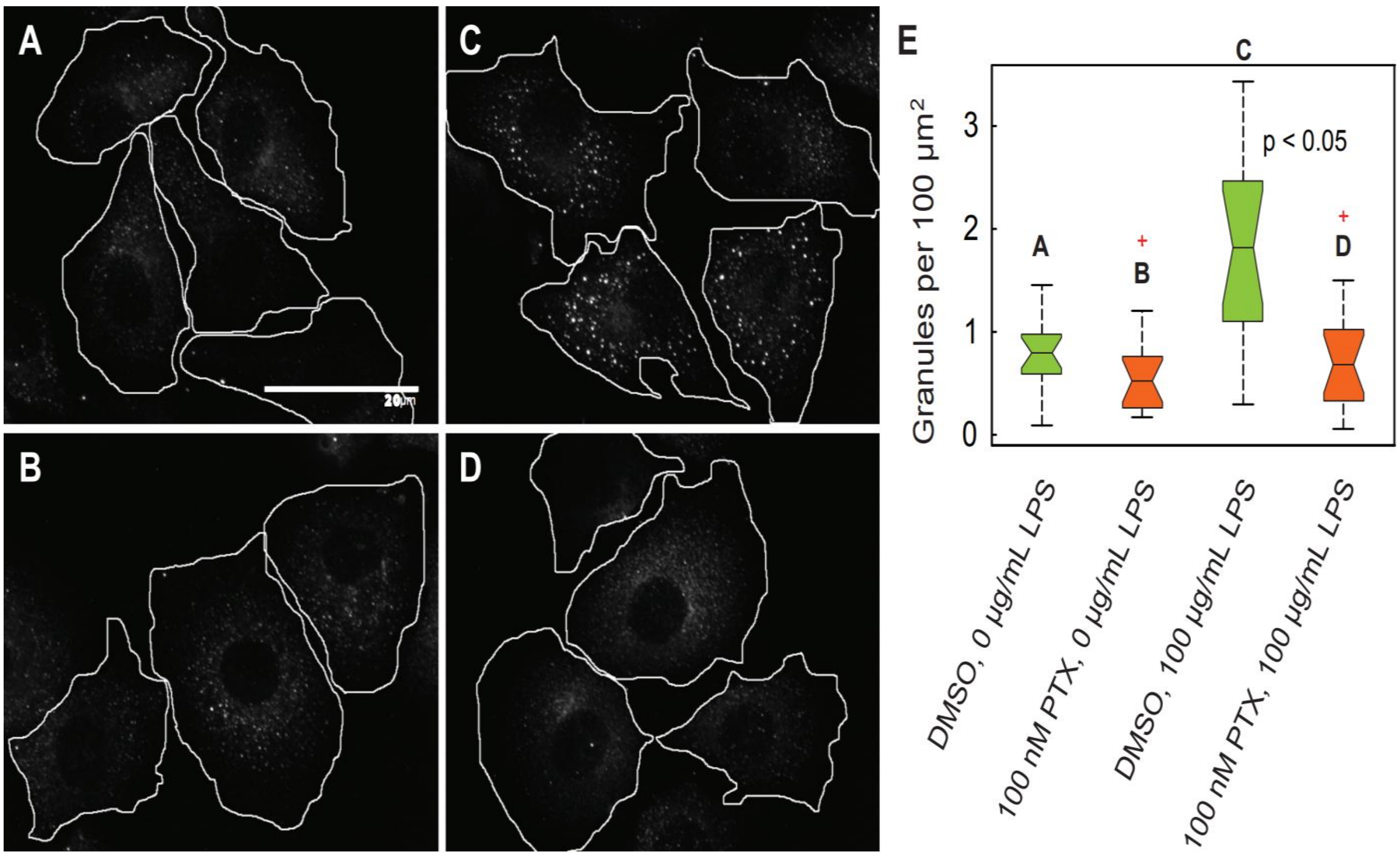

**Figure 3. Paclitaxel treatment prevented trafficking of intracellular TGFβ to the cell surface in MDA-MB-231 cells.** (A-D) White solid lines denote the segmented cell boundaries. Detection of TGFβ at the cell membrane by immunofluorescence in non-permeablized MDA-MB-231 cells stimulated with 100 μg/mL LPS stimulation followed by vehicle (DMSO) or 100 nM paclitaxel (PTX) treatment. See also the x-axis labels in (E). Three to five cells were analyzed per condition. Scale bar equals 20 μm. (E) Distributions of the number of TGFβ-stained foci on the cell surface.

## Sensitivity of diffuse gastric cancer cells to docetaxel

Although taxane-based therapy is approved for advanced gastric cancer, a majority of patients exhibit intrinsic resistance to taxanes and require additional lines of therapy (Shitara et al., 2020). Identifying methods to predict drug resistance would spare the individuals predisposed not to respond to taxanes from the side effects. Intestinal and diffuse gastric cancer are the two major intrinsic genomic subtypes with distinct patterns of gene expression. Intestinal cell lines are significantly more sensitive to 5-fluorouracil and oxaliplatin, but more resistant to cisplatin than the diffuse cell lines (Tan et al., 2011), which carry *TP53* mutations. In this context, to investigate the sensitivity to taxanes in diffuse gastric cancer, we performed titration experiments with eight diffuse gastric cancer cell lines and compared their sensitivity to docetaxel. In all three sensitive cell lines (TMK1, SNU1, and AZ521), the cell survival was less than 50% for docetaxel concentrations of 10 nM and higher, and the MT network formed bundles (Fig. 4). Our results suggest the clinical use of a docetaxel and cisplatin regimen for gastric cancers with mutated *TP53*.

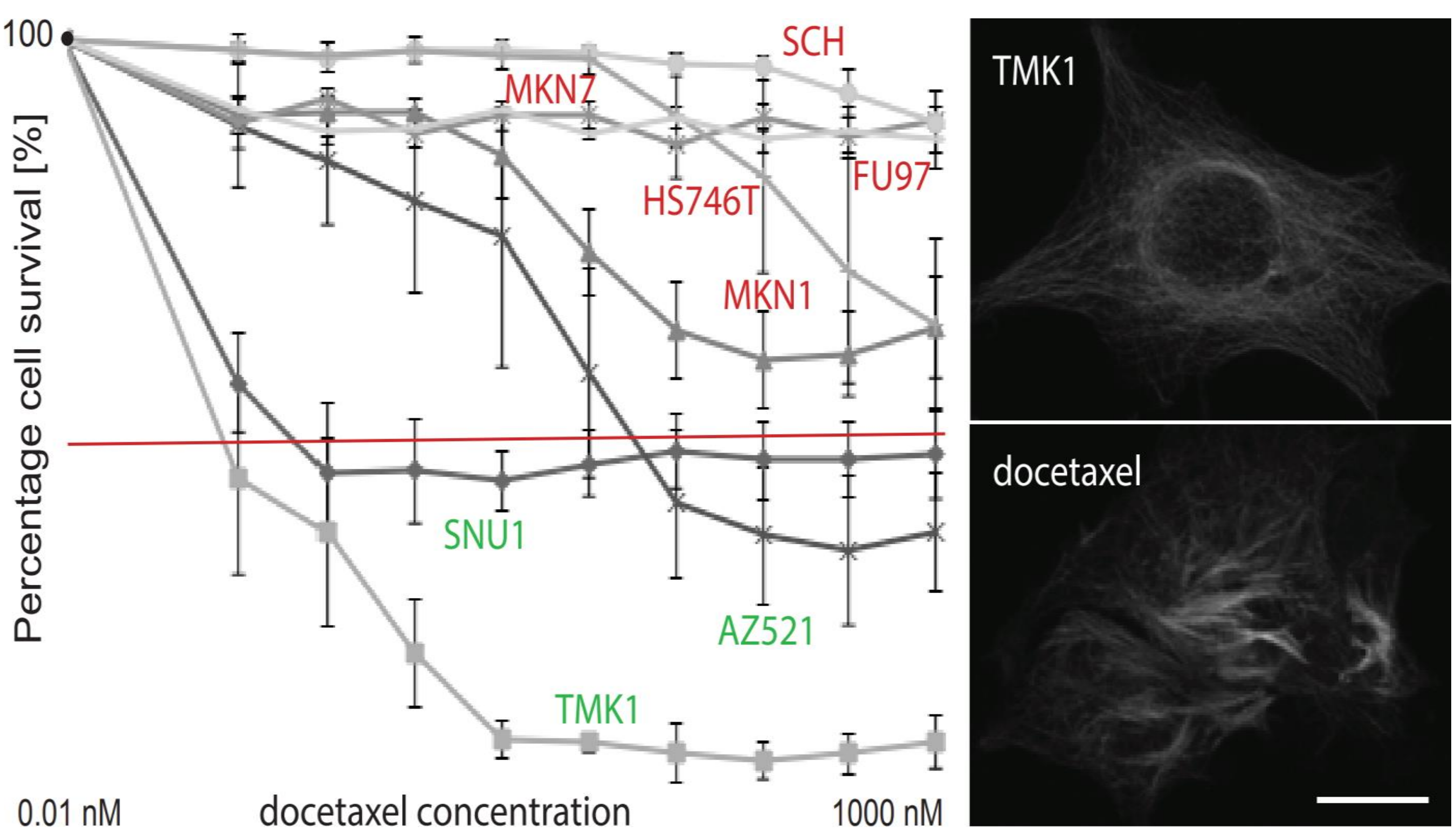

**Figure 4. Sensitivity to docetaxel in diffuse gastric cancer cell lines.** The figure shows cell survival curves for eight cell lines treated with increasing doses of docetaxel. The right panels show the MT cytoskeleton in a drug-sensitive TMK1 cell before (upper panel) and after (lower panel) treatment with 100 nM docetaxel. Scale bar equals 10 µm. Magnification 63x.

Cell proliferation dynamics follow a Gometz function (Gompertz, 1825; Winsor, 1932). Further, the growth of patient-derived organoids seeded from single cells could undergo a bifurcation point in gene expression (Bratsun et al., 2005; Gillespie, 2007) and either continue exponentially or undergo cell death (Matov, 2024h). After drug treatment, there is a lag in the consequential reduction in organoid size - resembling a hysteresis cycle (Ewing, 1881) - which can be utilized as a metric for drug efficacy evaluation. To model tumor growth dynamics, we will model organoid growth as a Markov chain with absorbing states and will use Bayes rules statistics (Bremaud, 1999; Matov, 1999).

**Enumeration of circulating tumor cells**

The molecular mechanisms underlying taxane resistance in prostate cancer have not been well elucidated due to the limitations in the available tumor tissue to study. CTCs represent a liquid biopsy of the tumors and CTC isolation can lead to molecular characterizations potentially revealing predictive biomarkers for taxane sensitivity or resistance (Yu et al., 2014). CTC can be isolated using microfluidic devices for capture and light-induced release of viable CTCs (LeValley et al., 2019; Liu et al., 2022). In the example we are presenting here, a ficolling technology is used to isolate live CTCs from whole blood of prostate cancer patients (Tasaki et al., 2014). These cells are plated onto an 8 mm coverslip (Matov, 2024e) and subsequently stained for three labels used for enumeration at low magnification – DAPI, to label the cell nucleus, PSMA is used as positive CTC identifier, CD45 labels the leukocytes and is therefore used as negative CTC identifier (Sung et al., 2013; Sung et al., 2012). Using these three labels, we aim at differentiating the prostate cells from white blood cells isolated from the blood's buffy coat.

To this end, we developed a computer vision algorithm for the automated analysis of blood samples (see Materials and Methods). In brief, it performs stationary wavelet transform-based segmentation to detect local intensity clusters through a combination of multiscale products (Starck et al., 2000) and denoising by iterative filtering of significant coefficients (Olivo-Marin, 2002). Our analysis enumerates the number of CTCs in patient blood and an example for one of the 50 positions (with 3 detected CTCs) resulting from imaging a coverslip at 10x magnification is shown in Fig. 5. We will aim at correlating this information with taxane efficacy in patients to develop a predictive algorithm that can identify taxane resistant patients before they receive treatment. Altogether, our computational approach connects exciting basic science with clinically relevant concepts and customization of chemotherapy.

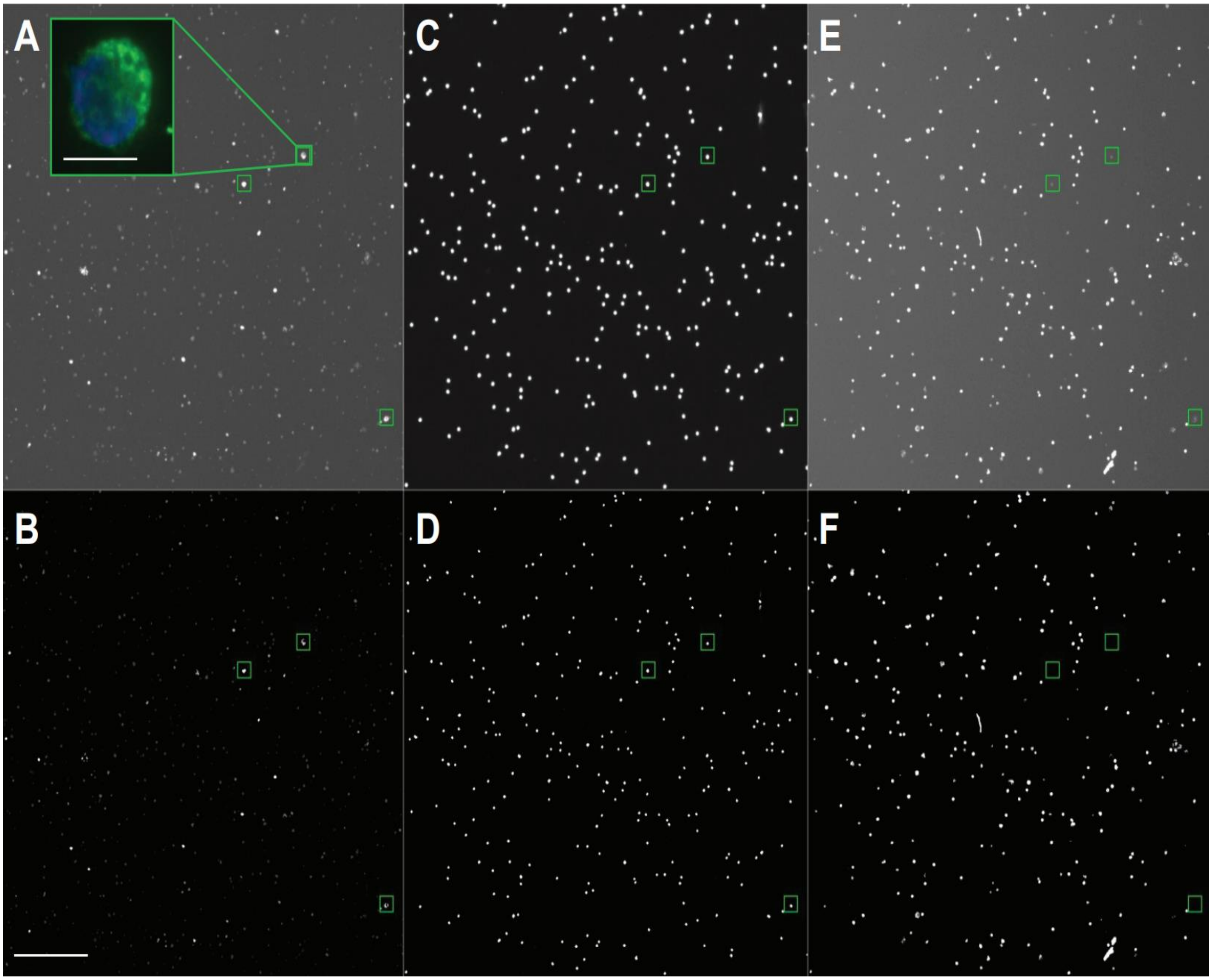

**Figure 5. Enumeration of patient CTCs in samples imaged on coverslips.** The figure (showing 1 of 50 images to cover the full area of the coverslip) demonstrates an approach for the detection of CTCs of metastatic prostate cancer patients using multiplex fluorescence microscopy. (A) PSMA labeling. (B) Segmentation of the PSMA image shown in (A). (C) DAPI labeling of nuclear areas. (D) Segmentation of nuclei shown in (C). (E) CD45 labeling of leukocytes. (F) Segmentation of the CD45 image shown in (E). The tumor cells are detected as PSMA+/DAP+/CD45- cells and marked with green squares. Scale bar equals 200 µm. Magnification 10x. Inset figure legend: PSMA (muJ591, tumor marker, green), DAPI (nuclear marker, blue). Scale bar equals 20 µm. Magnification 40x.

**Patient-derived organoids**

Patient-derived organoids are small organ-mimetics that better recapitulate, compared to cell lines, *in vivo* responses to treatments (Sachs and Clevers, 2014). Organoids can be derived from patient CTCs and tissue biopsies, and we worked with prostate and colorectal tumor samples. In our experience, lentiviral transduction allows for fluorescent markers to be introduced and imaged in patient cells forming organoids within 14 hours after surgery (Matov, 2024f; Matov, 2025a). Such clinically relevant cultures can be utilized to support decisions regarding the choice of regimens in the context of anticipation of drug resistance. In prostate cancer, tubulin inhibitors are widely used in advanced disease (Petrylak et al., 2004; Tannock et al., 2004) without having precise predictions regarding their clinical efficacy in terms of both intrinsic and acquired resistance.

Our hypothesis is that drug susceptibility can be evaluated *ex vivo* not only by direct analysis of patient tumor cells, but also based on the response of adjacent to the tumor tissues. In primary prostate cancer (Fig. 6), this approach can be implemented by culturing adjacent benign epithelial cells as well as cancer-associated fibroblasts (Matov, 2025b). In the metastatic setting, such evaluation can be accomplished also by analysis of cells in the host tissue. Cells from tissues of different origin, for example colorectal and lung, do not mix in organoids. We established organoids from both metastatic colorectal cancer cells and benign lung cells derived from a chest wall resection (Fig. 6). The availability of the cells from the tumor microenvironment, such as the extracellular matrix, can improve the precision of the analysis *ex vivo* (Matov, 2025a).

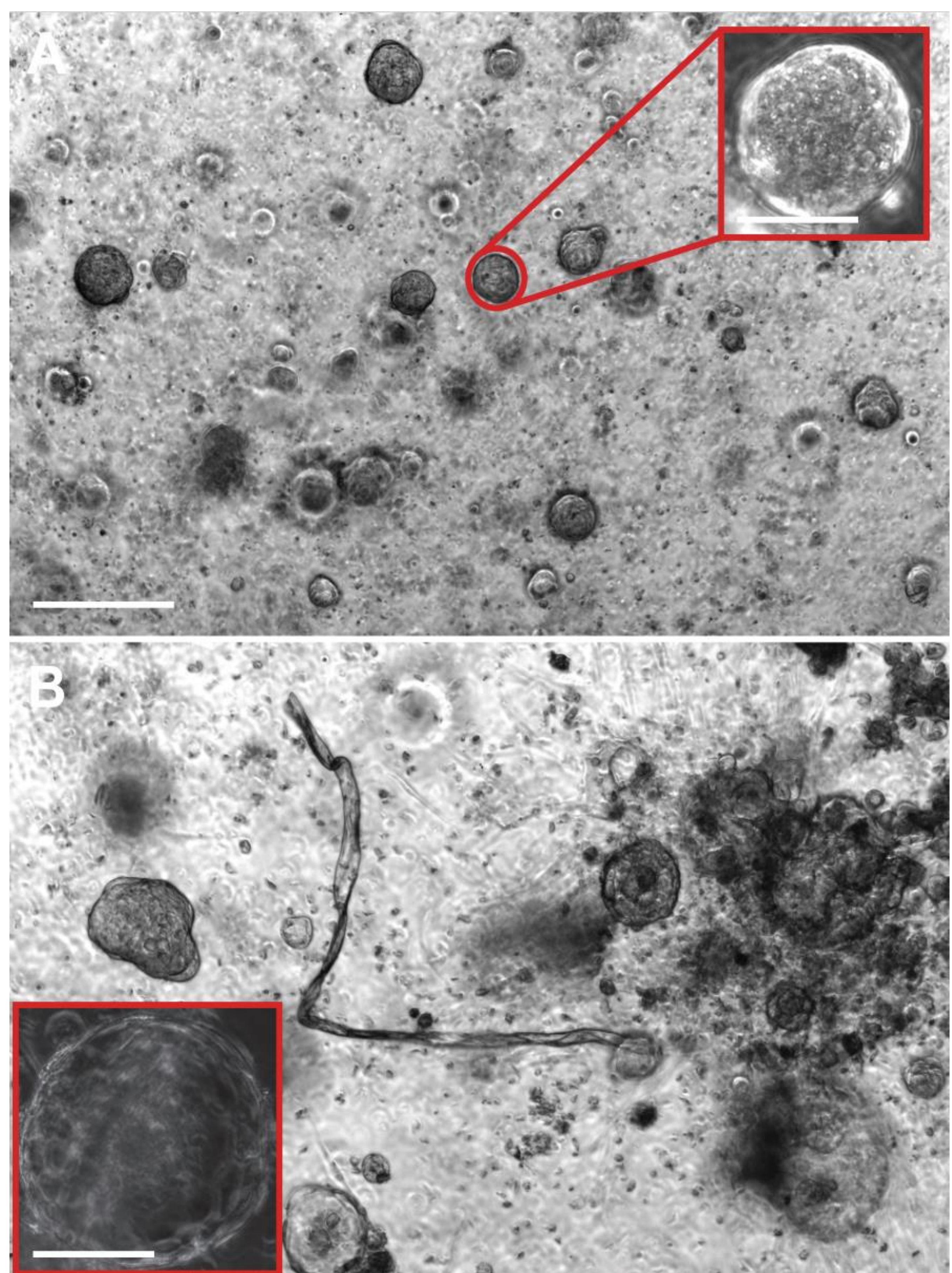

**Figure 6. Patient-derived organoids.** Transmitted light microscopy, magnification 4x. Scale bar equals 250 µm. Insets: phase contrast microscopy, 20x magnification. (A) Organoids derived from radical prostatectomy tissue from prostate tumor with Gleason Score 5+4. Day 13. Inset scale bar equals 50 µm. (B) Organoids derived from video-assisted thoracoscopic surgery of a metastatic rectal tumor protruding from the sternum. Day 25. Inset: Lung organoid. Inset scale bar equals 200 µm.

Because of the limited amount of tumor tissue collected during biopsies, it is not feasible to perform the traditional toxicity assays, such as $IC_{50}$ measurements. It is our hypothesis that assays related to killing tumor cells *ex vivo* do not faithfully represent the ability of the drug to cause patient tumors to recede. Furthermore, many primary prostate tumors are dormant and do not allow expansion *ex vivo*, and even in the case of proliferative cultures, with every passage the cells additionally diverge from the original tumor regulation. For these reasons, we propose the utilization of assays that demonstrate the engagement, by the drugs, of their cellular target without killing the organoids. Changes in protein morphology and dynamics in patient-derived cultures after drug treatment will, thus, become – upon clinical validation - metrics for drug susceptibility. This approach will improve our ability to anticipate drug resistance and spare patients that are not predisposed to respond to a certain drug the toxicity. In addition, it will allow us to identify an optimal drug regimen for each patient.

### Microtubule tip motion analysis

Fluorescent comets form at the polymerizing ends of MTs when the end binding protein 1/3 (EB1/3) – which recognizes the nucleotide state and GTP-bound tubulin (Zanic et al., 2009) – is GFP-labeled and allow for direct measurements of MT plus end, but not minus end, dynamics in living cells (Matov et al., 2010; Matov et al., 2006; Matov et al., 2005). Harvesting EB1 data allows for the generation of large, statistically reprehensive datasets in which for every cell we can measure over 38,000 data points (Fig. 7). In cell lines, we analyze 10 to 20 cells (that is, between ½ and 1 million quantitative readouts, see Vid. 2 and Vid. 3 for examples) per condition to evaluate drug susceptibility at very low drug doses (Shah et al., 2021) by measuring the specific patterns of changes in MT dynamics (Matov, 2024c;

Matov, 2024f). In patient-derived organoids, it is difficult to know the genetic and epigenetic make-up of the cells in an organoid, in particular in primary, drug-naïve tumors. We performed EB1 comet analysis in organoid cells (Vid. 4) from four metastatic castrate-resistant prostate tumors, which showed correlation between MT dynamics and gene expression of MT-regulators (Matov, 2025a), and our predictions regarding drug resistance can be validated in a clinical trial.

A plethora of the cellular proteins exhibit ambiguity (and redundancy) in function and response to perturbations. There exists a high level of regional variability in the dynamics of polymer networks in living cells, which are considerably more interdigitated than initially perceived and reported in the literature (Matov, 2024d). Drugs bind to proteins within these networks and affect their function. Consequentially, the modulation (Matov, 2024g) of MT and F-actin dynamics inevitably affects drug efficacy. Answers about the potential efficacy of a drug can, thus, be obtained by utilizing clinical software systems that measure, in great detail, the effects of a regimen on its cellular targets (Matov, 2024h). Molecular manipulations of living patient-derived cells *ex vivo* can reveal which secondary mechanisms would be activated after a particular intervention.

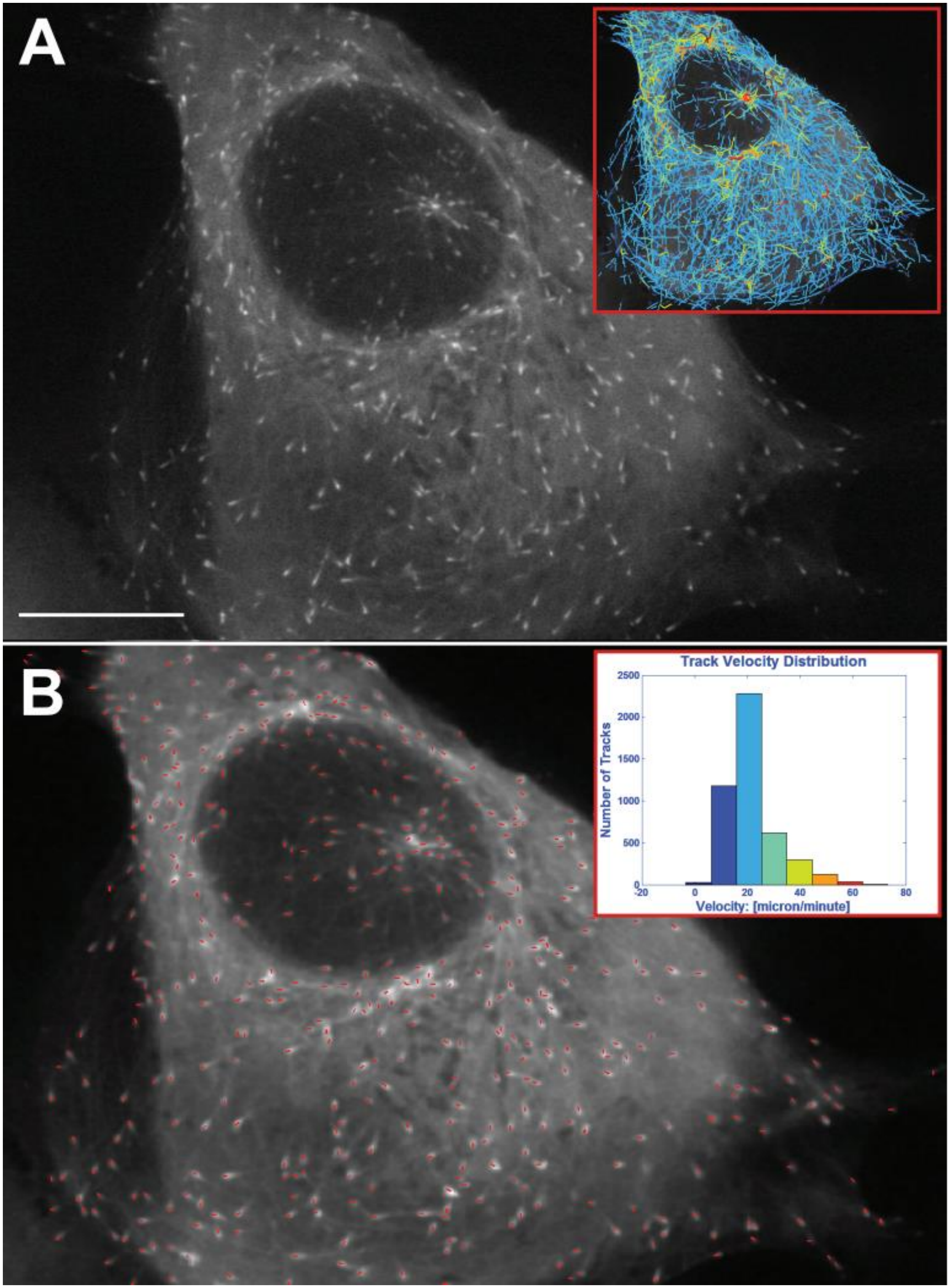

**Figure 7. EB1 comets detection and motion tracking in MDA-MB-231 breast cancer cells.** (A) MT tips are labeled with EB1ΔC-2xEGFP and imaged for a minute with an acquisition rate of two images per second. (B) EB1 comets formed at the tips of polymerizing MTs are detected (Matov et al., 2011) and computationally tracked (Yang et al., 2005). The upper inset displays an overlay with the EB1 trajectories for comets persisting in at least four image frames, that is, at least 1.5 seconds. The lower inset shows the histogram of the EB1 speed probability density function of these trajectories, which comprise over 38,000 comets. The average speed is 24.4 ± 9.4 μm/min. The maximal EB1 speed is over 70 μm/min. The speed distribution is unimodal. The color-coding represents EB1 speeds and colder colors correspond to lower speeds, and warmer colors correspond to faster speeds. Scale bar equals 10 μm.

We have shown that the rates of MT polymerization correlate with the size of the GTP-cap at MT plus ends – and that EB3 comets become less eccentric as they approach the cell edge (Matov et al., 2005) and their speeds are reduced on average with 3.2 μm/min (Matov et al., 2006). EB1 binding accelerates conformational maturation in the MT, most likely by promoting lateral protofilament interactions and by accelerating reactions of the GTP hydrolysis cycle (Maurer et al., 2014). Our computational approach can also detect and measure changes in MT polymerization dynamics by analyzing other proteins that decorate the MT plus ends, such as GFP-labeled CLIP170 (Matov et al., 2010). CLIP170, like EB1/3, treadmills on polymerizing MT plus ends by binding GPT-capped tubulin and unbinding after hydrolysis to GDP (Perez et al., 1999). When spindle MTs simultaneously undergo polymerization and are also being translocated by molecular motors, the speeds we record for MT tips movement overestimate the actual turnover rate. Two-color time-lapse series (Matov et al., 2005) with EB1ΔC-2xEGFP and X-rhodamine labeled tubulin speckles (Ponti et al., 2003) allow us to decouple MT polymerization from polymer flux (sliding) by molecular motors (Matov, 2024g).

We have developed a computational assay for the quantification of 12 parameters of MT dynamics based on the analysis of EB1/3 comets (Matov et al., 2010). It is our hypothesis that the pattern of MT dynamics a tumor generates can be matched (Matov, 2006; Matov, 2011), like in a game of Tetris (Matov, 2025a), to a tubulin inhibitor (Matov, 2024e) affecting the same parameters. Our work has indicated that a dozen MT-regulating genes lead to changes in MT polymerization dynamics (Matov, 2025a) based on their differential regulation (Gao et al., 2014) in prostate cancer. The genes that regulate MT homeostasis have been shown to be implicated in conferring resistance to tubulin inhibitors (Kavallaris, 2010). Recent work has also linked them to resistance to RAS-targeting therapy.

For instance, TBK1 inhibition has been shown to reverse therapeutic resistance to sotorasib and adagrasib in colorectal cancer (Alonso et al., 2025). TBK1 localizes to centrosomes during mitosis and regulates MT dynamics. Depletion of TBK1 attenuates MT catastrophe frequencies and protects them from the depolymerizing activity of nocodazole, while the overexpression of TBK1 attenuates MT polymerization rates (Pillai et al., 2015). TBK1 binds to the centrosomal protein CEP170 and to the mitotic apparatus protein NuMA, and both CEP170 and NuMA are TBK1 substrates (Pillai et al., 2015). NuMA activates dynein and their activity organizes MT minus ends into focused spindle poles (Colombo et al., 2025; Merdes et al., 1996). Further, NuMA caps MT minus ends and attenuates MT catastrophe frequencies (Colombo et al., 2025). Selective disruption of the TBK1-CEP170 complex hyperstabilizes MTs and triggers defects in mitosis, suggesting that TBK1 functions as a mitotic kinase necessary for MT dynamicity and spindle organization (Pillai et al., 2015). Quantitative live-cell image analysis of MT dynamics in patient-derived cells can, thus, be utilized for refinement of combination therapies with $RAS^{G12C}$ inhibitors.

Such clinical research (Matov, 2006) can aid the discovery of an optimal drug selection (Matov, 2011) for each disease, which is the motivation to contribute to translating all image analysis of molecular markers from off-line software to real-time algorithms.

**Validation of drug susceptibility in patient cells**

In prostate cancer organoids derived from intraductal carcinoma, an aggressive type of adenocarcinoma (Dube et al., 1973; Rhamy et al., 1972) with very particular glandular organoids in primary samples (Matov, 2024g), we measured attenuated MT polymerization rates in comparison to other metastatic tumors (Matov, 2025a). In the intraductal carcinoma organoids, the MT-associated protein 2 (MAP2) was upregulated up to 23-fold in comparison to the other tumors. MAP2/tau stabilizes MTs by attenuating the polymerization rates by binding along the lattice (Al-Bassam et al., 2002), which are the changes we directly measured with our computer vision assay (Matov, 2025a). Such datasets can be validated with patient samples, such as urinary microRNAs (miRs) (Matov, 2024a; Matov, 2024j).

Non-invasive, for instance from urine samples, or minimally invasive methods for obtaining patient cells allow us to monitor disease progression and patient response to treatment throughout the course of a clinical trial. We can transduce patient cells and label proteins of interest overnight and, depending on the disease and drugs used, drug susceptibility testing can be carried out as soon as a few hours after sample collection. Having the ability to test a number of drugs and combinations *ex vivo* will be critical in the treatment of pathologies for which there is no known cure. Once the specific impaired molecular mechanisms is identified for the particular patient, treatment options that correct the aberrations can be selected. This way, we can transform patient treatment.

We analyzed urine samples from 15 healthy individuals and 13 drug-naïve stage IV lung cancer patients and detected 947 miRs, 20 of which could be used to discriminate healthy from lung cancer samples (Matov, 2024j). The biomarkers we detected are implicated in cellular functions, such as cell proliferation, which is directly linked to MT dynamics. miR-577 inhibits cell proliferation and invasion in non-small cell lung cancer (NSCLC) (Men et al., 2019). miR-141-3p reduces pulmonary hypoxia and reoxygenation injury (Zhan et al., 2024). miR-29c-3p is significantly decreased in the plasma of lung cancer patients and can be used as a biomarker discriminating between NSCLC and small cell lung cancer (SCLC) (Zhang et al., 2023). miR-95-3p inhibits the invasiveness of metastatic lung cancer through downregulation of cyclin D1 (Hwang et al., 2015). miR-335-5p is significantly decreased in parenchymal lung fibroblasts of smokers (Ong et al., 2019). miR-29a-3p prevents NSCLC tumor growth and cell proliferation, migration, and invasion by inhibiting the Wnt/β-catenin signaling pathway (Zhang et al., 2022). miR-532-3p inhibits metastasis and proliferation of NSCLC by targeting FOXP3 (Ni et al., 2021). There are other examples of urinary biomarkers that are validated in data from primary tumor tissues and plasma samples. Overall, the utilization of longitudinal urine samples will allow a clinical validation of predictions regarding drug treatment established in organoids by analysis of exosomes (Matov, 2024a) and the miRs they contain (Fig. 8).

Computational analyses of gene expression of MT-regulators in patient cells can provide hypotheses regarding drug efficacy, which can be validated by the combined utilization of live-cell computer vision and sequencing of miRs. This three-pronged approach will allow us to select an effective drug regimen, for example in immunotherapy. The MTs are involved in the formation of the immunological synapse (Editorial, 2004; Lagrue et al., 2013; Serrador et al., 2004) and taxane therapy affects the immune system (Serpico et al., 2020), and tubulin inhibitors might synergize with the immune cells under the right conditions (Fong et al., 2019). The links between systemic therapy with tubulin inhibitors and immunotherapy (Chamuleau et al., 2023; Dunleavy et al., 2018) in the context of the regulation of MTs and other cytoskeletal proteins (Ben-Shmuel et al., 2021; Burkhardt, 2013) seem of key importance for the success of immune cells in killing tumor cells of epithelial origin. For these reasons, we propose computational assays to anticipate drug resistance and facilitate the selection of an optimal drug regimen in oncology.

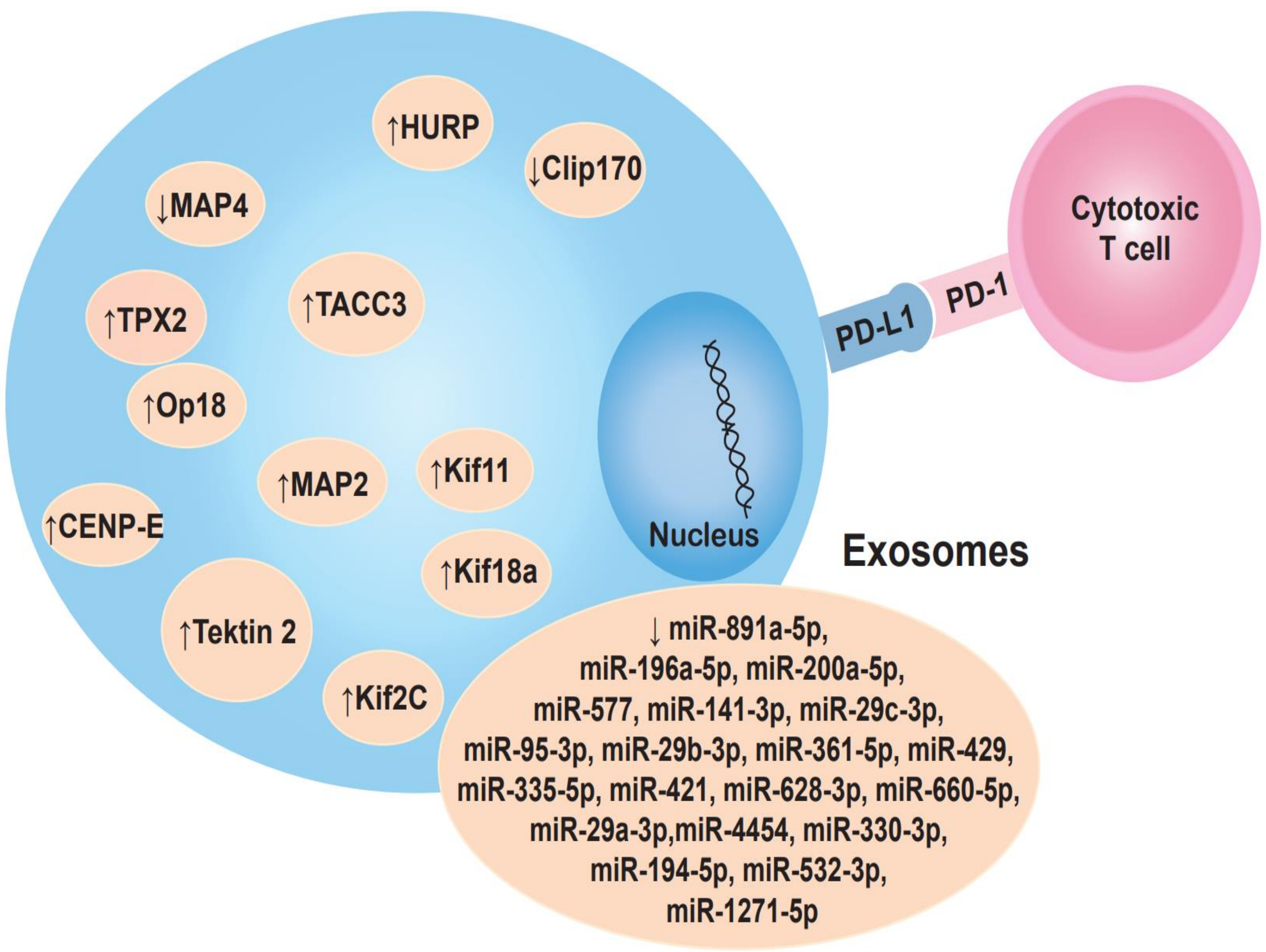

**Figure 8. Dysregulation of MT-regulating genes and miRs abundance levels in patient samples.** The figure presents novel urinary miR biomarkers in lung cancer (Matov, 2024g; Matov, 2024j) and a pattern of dysregulated MT-regulating genes we identified in intraductal prostate cancer (Matov, 2025a). In addition to next-generation sequencing and gene expression analysis, computer vision analysis of exosomes (see Vid. 5) can validate the drug resistance mechanisms.

## Microtubule dynamics regulation of immune response

EB1 interacts with T-cell receptor cytosolic regions and mediates the organization of an immunological synapse to transduce activation signals (Lasserre and Alcover, 2012). Lytic granules from cytotoxic T cells have been shown to exhibit kinesin-dependent motility on MTs (Burkhardt et al., 1993). T-cell activation and function can, therefore, be evaluated in terms of the reorganization and modulation of the cytoskeleton (Matov, 2024d; Matov, 2024g; Matov, 2024h; Matov and Danuser, 2004; Ponti et al., 2003) and the EB1 comet metrics will be validated by the levels of measured cytokine production. After validation and calibration, the software can be utilized for the discovery of novel drug combinations that allow personalized treatment.

We derived a prostate cancer organoid from micro-metastasis at a retroperitoneal lymph node for which whole genome sequencing showed homozygous mutations in *PTEN* and *BRCA2* as well as mutated KIF4B (Matov, 2024g). When prostate tumors escape the capsule, they form micro-metastatic lesions at the retroperitoneal lymph nodes (Brassetti et al., 2019; McLaughlin et al., 1976). In our organoid culture, such drug-naïve tumors formed slow-proliferating tumors and had driver mutations, such as homozygous mutations in *PTEN* (exon2, c.G194>C, p.C65S) and *BRCA2* (exon14, c.T7397>C, p.V2466A) (Matov, 2025b). A heterozygous mutation we identified in *KIF4B* (exon1, c.G1739>T, p.R580L) likely contributes to errors during chromosome segregation (Mazumdar et al., 2004). We detected also androgen receptor (AR) deletion, which we have shown to be associated with smaller tumor cells and more aggressive disease (Matov, 2024c; Matov, 2024e). Overall, these are examples for which image analysis in living cells can validate the mechanisms suggested by sequencing data and, thus, inform therapy.

Cell therapy (Gage, 1998) is becoming the forefront of precision medicine, and it would be critical to be able to anticipate the mechanisms of action of the engineered immune cells (Dekkers et al., 2023; Irvine et al., 2022; Lim and June, 2017). Natural killer cells and cytotoxic T cells employ different mechanisms to kill their targets (Deng et al., 2015; Mitchison, 2021; Pardoll, 2003; Pardoll and Topalian, 1998), for instance by secreting cytotoxic lysosomes using the MT cytoskeleton for trafficking and release. There is also a plethora of effects induced in the target cells, most well-studied of which lead to apoptosis and ferroptosis (Dixon et al., 2012; Viswanathan et al., 2017), a recently identified potent, caspase-independent, mechanism of programmed cell death. The exact pathways activated during therapy can be pinpointed by measuring its effects on the target proteins in patient-derived living cells *ex vivo*.

It is conceivable that for different patients, with overall very similar genetic profiles, a distinct type of engineered immune cells will be required. Given the availability of high-quality fundamental research equipment for high resolution live-cell microscopy in most university hospitals, what we propose is to embed in these systems such real-time software for automated quantification and on-the-fly statistical analysis of intracellular behavior. Introducing this quantitative imaging method to the clinic will allow physicians to fine-tune, with a high level of certainty, an optimal treatment regimen for each patient.

**Analysis of fundus fluorescence**

Neurodegenerative diseases are challenging to diagnose and stage. As an extension of the central nervous system (De Groef and Cordeiro, 2018), the eye harbors retina ganglion cells vulnerable to degeneration, and the fundus exhibits symptoms of early manifestations of amyloid protein aggregation linked to neurodegeneration (Vujosevic et al., 2023). The diagnosis of patients having a disease associated with protein aggregation could be based on capturing fundus auto-fluorescence (AF) images of the retina, but such baseline images have low contrast and low signal-to-noise ratio. We performed analysis in fundus AF image datasets obtained from non-human primates with automated detection algorithms, such as difference of Gaussians (Matov, 2024b) and stationary wavelet transform (Olivo-Marin, 2002). The largest baseline AF puncta (Fig. 9) detected (or segmented) had a peak intensity of 15.7% of the maximal intensity in the image and a perimeter (length of the boundary) of 200 μm, which indicates a utility of a methodology for enhanced fluorescence emission.

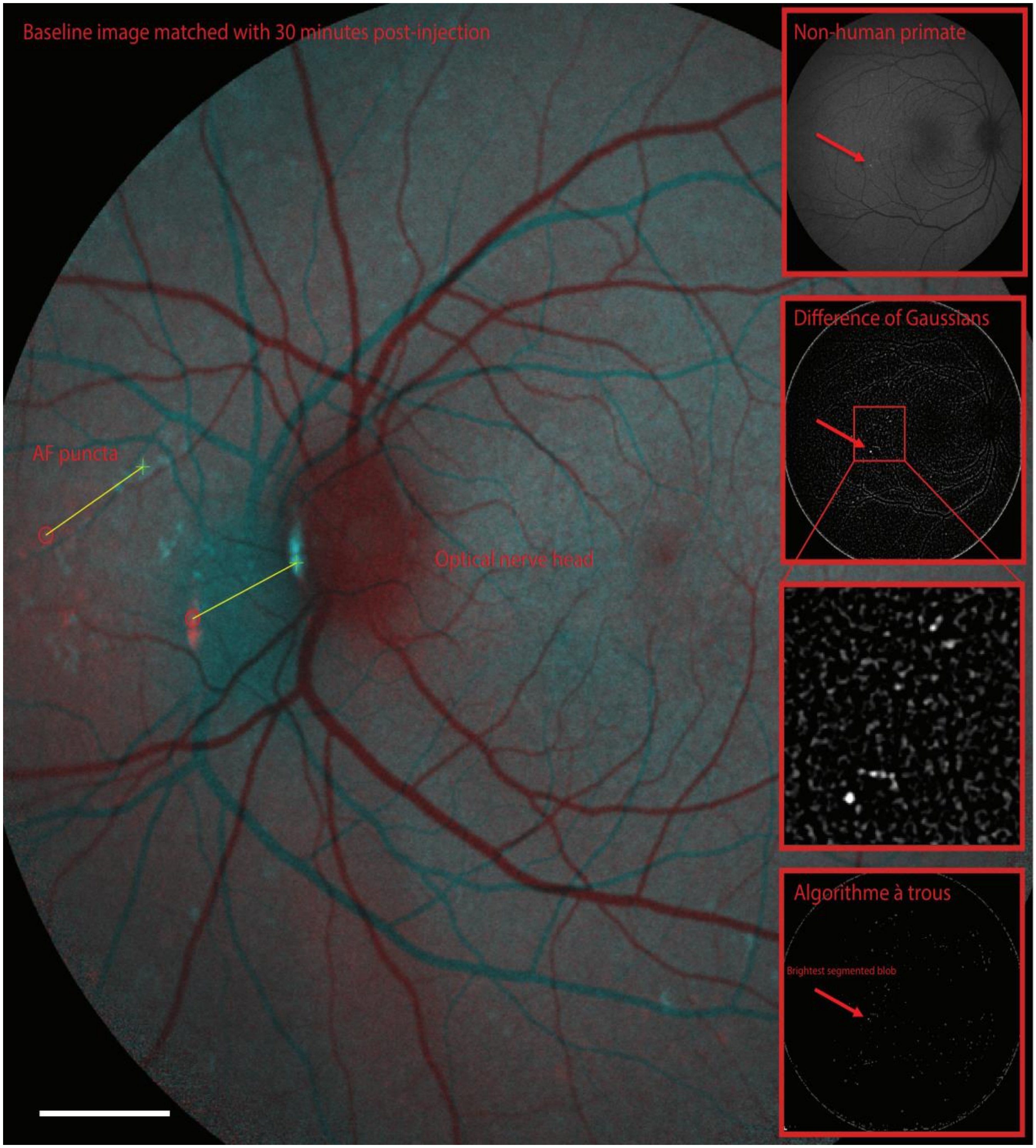

**Figure 9. Eye fundus with matched auto-fluorescence at baseline and hyper-fluorescent puncta 30 minutes post intravenous injection of a fluorescence emitting dye.** *In vivo* imaging of fundus of the eye in two time-frames, including prior to injection and 30 minutes post injection (Pilotte et al., 2024). AF and hyper-fluorescent puncta are indicative of disease. We show matching of two puncta, one in the area of the optical nerve head and another in the retina, which indicates neurodegeneration. Matching was done using speeded-up robust features (SURF) (Bay et al., 2008). Inset images show another example of fundus fluorescence imaging and the detection of AF puncta with a band-pass filter (difference of Gaussians) (Matov et al., 2011) and stationary wavelet transform (algorithme à trous) (Olivo-Marin, 2002). The automatically selected area was the brightest segmented blob after image transformation. Scale bar equals 2 mm.

An amyloid-binding fluorophore can enhance fundus imaging (Aguilar-Calvo et al., 2022; Pilotte et al., 2024; Pilotte et al., 2025) and aid the analysis of amyloid deposits. Computer visions analysis of compounds that emit fundus fluorescence upon binding to a misfolded or aggregated protein can facilitate the diagnosis of Alzheimer's disease, cerebral amyloid angiopathy, traumatic brain injury, glaucoma, age-related macular degeneration, Parkinson's disease, multiple system atrophy, dementia with Lewy bodies, amyotrophic lateral sclerosis, and a dozen others. Immediately after injection, the dye enters the retina (about 15 seconds post-injection) and fills the retinal vasculature rapidly. Figure 9 shows longitudinal matching of two AF puncta, one in the area of the optical nerve head and another in the retina, with hyper-fluorescent signals 30 minutes after injection with a fluorescent emitting dye binding to misfolded or aggregated proteins by applying a SURF detector (Bay et al., 2008) (see Materials and Methods). The *in vivo* analysis of fundus fluorescence also faces a challenge in the continuous 3D rotation of the eye during image acquisition and requires image registration (Bajscy and Broit, 1982; Barrow et al., 1977). In this context, a real-time image registration and segmentation software will facilitate the analysis of fundus fluorescence puncta.

**DataSet Tracker**

Undoubtedly, all degenerative diseases are associated with impairment in intracellular trafficking. The highly dynamic organization and remodeling of the cellular cytoskeleton are dysfunctional in pathology and often lead to drug resistance. Successful analyses of the mechanism of drug action require statistical analysis of large-scale readouts of the changes in molecular interactions. Our objective has been to develop resources for functional interrogation of drug response in a physiologically relevant system amenable to molecular manipulations and investigate personalized drug response *ex vivo*. We have developed image analysis software for automated motion tracking of labeled MTs and F-actin – ClusterTrack (Matov et al., 2010) (for measurements in interphase cells) and Instantaneous Flow Tracking Algorithm (IFTA) (Matov, 2024b; Matov et al., 2011) (for measurements of interdigitated flows in dividing cells, contractile F-actin meshworks in migrating epithelial cells and neuronal growth cones) – and other biomedical computer vision software packages (Matov, 2024h). We used IFTA to constrain the Kalman filter (Cameron et al., 2006; Yang et al., 2008; Yang et al., 2007; Yang et al., 2005) with information of spatio-temporal gradients in the instantaneous flow behaviors, which increased the Kalman tracking success rate – which is the number of correct tracks divided by the overall number of tracks – from 61% to 96% (Fig. 10).

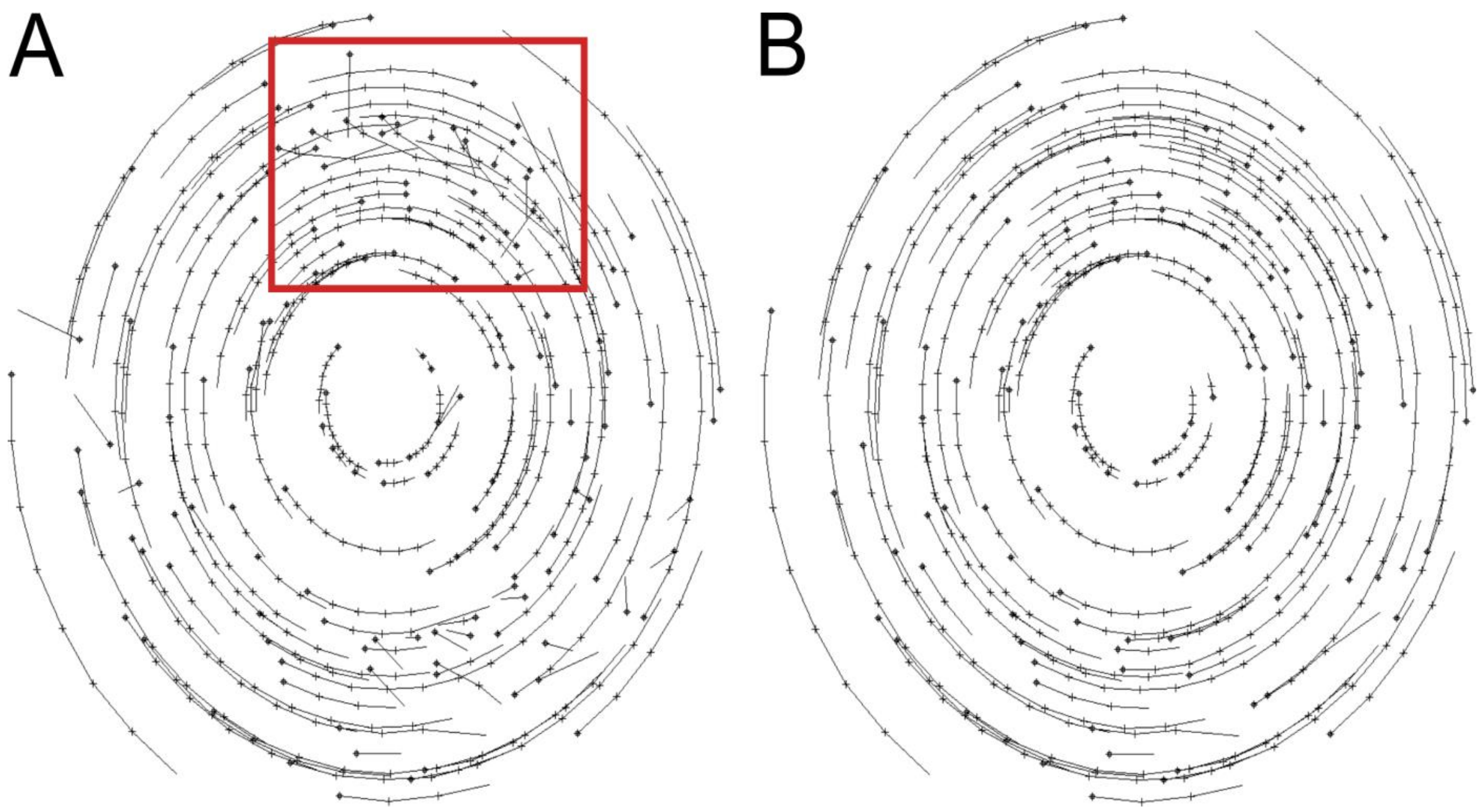

**Figure 10. Tracking the motion of synthetic markers in time-lapse image series.** The figure (adapted from (Yang et al., 2005)) presents the changes in Kalman filtering tracking results when a flow constraint based on IFTA is imposed during tracking in the rotating dish sequence, consisting of 10 frames with 80 features each. Figure legend: A square marks a trajectory starting point, a cross marks the selected for the trajectory feature in the next time-frame. (A) Tracking results without IFTA flow constraint. Success rate: 61.1%. Red square highlights an area with multiple wrong links. (B) Tracking results with flow constraint. Imposing IFTA flow constraints improves the tracking success rate to 96.0%.

All quantitative imaging examples presented here will benefit from having the analysis results displayed on the microscope screen on-the-fly during sample observation and image acquisition. In this context, the manuscript explores, for the first time, a real-time quantitative way to analyze the changes occurring in a cell. In an effort to improve the efficiency in drug development and efficacy evaluation, we have developed a real-time computer vision software, which connects to the microscope camera, processes multiple image frames per second (5 frames per second for the example in Fig. 11 – see the lower right corner of the image and Vid. 6), and instantly displays and stores statistical readouts in parallel to sample observation and image acquisition.

Our objective is to offer a tool not only capable to anticipate drug resistance for existing regimens, but methodology that can also be seamlessly applied in the identification of putative molecular targets during drug discovery. Quantifying with ease the morphology, localization, and other metrics of the dynamics of proteins in patient cells can certainly help clinical decisions. It can support timely changes in the original course of treatment by providing feedback and evidence for treatment efficacy, or the lack thereof, in real time. It will also advance our understanding of the regulation of the cells comprising our organs and tissues in normal physiology and in disease.

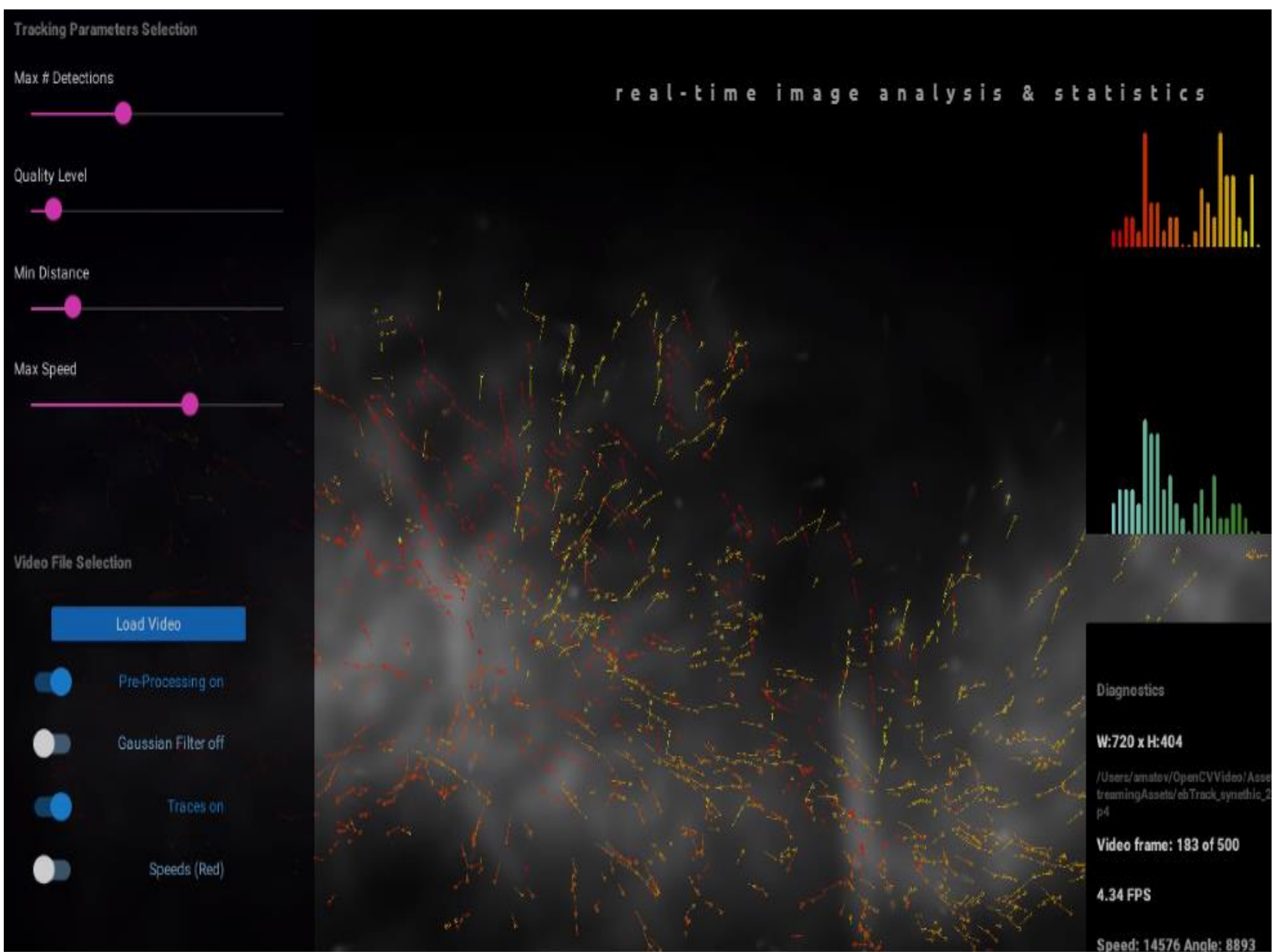

**Figure 11. Real-time tracking the motion of synthetic markers in time-lapse image series.** The figure presents novel software (a demo video of this package is available for viewing in Vid. 1) developed with the gaming engine of Unity Technologies. The figure shows computer vision analysis of the motion of synthetic markers, which mimic live-cell fluorescent microscopy image sequences. Displacement vectors color-coding is used to show the angular direction as well as the speed of motion. A button selection allows changing the display preferences. In the figure, yellow vectors move to the right and are also shown in yellow within the right peak of the bi-modal histogram in the upper right corner of the screen. Similarly, the vectors moving to the left are in red, both on the image overlay and within the left peak of the bi-modal histogram in the upper right corner of the screen. The second display option (not shown) changes the displacement vectors color-coding to showing different shades of green, depending on the speed. Observe in the unimodal histogram to the right that most of the features move slowly (the light-green peak to the left), while a few features move very fast (the dark-green distribution tail on the right side). Real-time information on the frames per second analyzed, the average values for the speed, and the angular vector orientations are displayed in the lower right corner of the screen. On the left side of the screen, there are slider buttons in the upper left corner, which allow to set the (i) the upper limit for the number of detected features based on the *a priori* knowledge of the nature of the motion in the analyzed sample, (ii) the level of statistical significance for the feature selection step, that is, the level of feature detection stringency, (iii) the minimum distance between features, which is another parameter selection done based on *a priori* knowledge of the type of sample analyzed, and (iv) a cut-off for the feature search radius, which limits the maximal allowed displacement; this is another parameter, which is selected based on *a priori* knowledge of the sample. By providing sample-specific input to the tracking module, the parameters selection allows to limit the computational complexity, to minimize the tracking errors, and to deliver the fastest analysis results. The blue buttons in the lower-left corner of the screen allow to change various aspects of the screen display in terms of showing image segmentation or motion tracking results, single-segment tracks (between just two frames) or the aggregated trajectories and, as described above, the vector color-coding (angles in red/yellow vs. speeds in different shades of green). See Materials and Methods for information on how to download and test the software. We will extend the current real-time 2D functionality to 3D analysis using AI algorithms.

## Algorithm for real-time optical tracking

We developed our novel software package (a demo video of this package is available for viewing in Vid. 1) with the gaming engine Unity (Unity, 2017). The engine supports desktop, mobile, console, augmented reality, and virtual reality platforms. It is convenient for iOS and Android mobile applications development. We will extend the current real-time 2D functionality to 3D analysis using AI algorithms.

Figure 11 shows computer vision analysis of the motion of synthetic markers, which mimic live-cell fluorescent microscopy image sequences. Displacement vectors color-coding is used to show the angular direction as well as the speed of motion. A button selection allows changing the display preferences. We have displayed a color-coding that gradually changes the color from yellow to red as vectors change their angular orientation from $90^o$ to the right to $90^o$ to the left, that is, features moving upwards are color-coded in orange, accordingly. In this example, yellow vectors that move to the right and are shown as the right peak of the distribution in yellow of the bi-modal histogram. Similarly, the vectors moving to the left are in red, both on the image overlay and within the left peak of the bi-modal histogram in the upper right corner of the screen. The second display option (only shown as a histogram) changes the displacement vectors color-coding to showing different shades of green, depending on the speed, with slow moving features are shown in light blue and the color is gradually changing toward green as the speed increases (right side of the histogram). Observe in the unimodal histogram to the right that most of the features exhibit slower speeds (the light-green peak to the left), while a few features move very fast (the dark-green distribution tail on the right side). Real-time information on the frames per second (5 frames/second) analyzed, the average values for the speed, and the angular vector orientations are displayed in the lower right corner of the screen.

On the left side of the screen, there are slider buttons for the selection of analysis parameters (in the upper left corner), which allow to set the (i) the upper limit for the number of detected features based on the *a priori* knowledge of the analyzed sample, (ii) the level of statistical significance for the feature selection step, that is, the level of feature statistical detection stringency, (iii) the minimum distance between features, which is another parameter selection done based on *a priori* knowledge of the type of sample analyzed, and (iv) a cut-off for the feature search radius, which limits the maximal allowed displacement; this is a parameter, which is selected based on *a priori* knowledge of the limits in the speed of motion in the analyzed sample. By providing sample-specific input to the tracking module, the parameters selection allows us to limit the computational complexity, to minimize the tracking errors, and to deliver the fastest analysis results. The blue buttons in the lower-left corner of the screen allow to change various aspects of the screen display in terms of showing image segmentation or motion tracking results, single-segment tracks (between just two frames) or the aggregated trajectories as the video progresses and, as described above, the vector color-coding (angles in red/yellow vs. speeds in different shades of blue/green).

For the purposes of demonstrating the workings of our real-time analysis software, we have generated synthetic videos displaying the radial motion of fluorescent features originating from two distinct focal points (Fig. 11, Vid. 6). In this contribution, we pre-process images making up the video and select features for motion tracking on-the-fly as the video file is loaded into the software. For each feature type, we will extract a set of distinctive key features, identify descriptor vectors of the features, and compute motion metrics to distinguish between drug-resistant and drug-sensitive profiles in cells derived from patients and elucidate mechanisms of resistance. This quantitative analysis (Vid. 7, Vid. 8) will, therefore, allow for the identification of the key signaling pathways involved in drug resistance, which may contribute to the personalization of the treatment regimens and the functional testing of novel compounds. Our approach will allow evaluating the relative contribution of different signaling pathways in drug response in disease cells originating from different organs and tissues.

Our preliminary software development uses a 2D tracking strategy in which we utilize three steps performed by well-established algorithms implemented as real-time robotics libraries by OpenCV (Bradski, 2000; Pulli et al., 2012). In brief, images making up the video are pre-processed or de-noised by a specialized background subtraction method (Zivkovic, 2004; Zivkovic, 2006). The algorithm aggregates statistical representation of the intensity profile for each pixel over time during live camera operation by the use of Gaussian (de Moivre, 1718) mixture models (Duda and Hart, 1973) in which the number of Gaussian function components is recursively updated with every new image. Lagrange multipliers (Lagrange, 1788) are introduced in the estimation of the maximum likelihood (Fisher, 1912) to determine the number of components. The influence of previous images in computing the distributions for each pixel is exponentially decaying with time. For multinomial distributions, a conjugate prior is utilized based on a Dirichlet distribution (Dirichlet, 1839) related to the minimum message length criterion (Wallace and Boulton, 1968).

Next, a watershed-based algorithm was used to select the features for tracking (Shi and Tomasi, 1994). The algorithm monitors the quality of the features during tracking by estimating feature dissimilarity between the first and current frames. The section procedure evaluates feature texturedness as it relates to the tracking accuracy. To optimize the tracking accuracy, metrics for two motion models for affine image changes (linear warping) and translation, are computed by Newton-Raphson (Wallis, 1685) minimization. Translation is the better metric when the inter-frame camera translation is small, while affine changes are important for determining dissimilarity between distant frames. The method successfully detects feature occlusions and disocclusions, and selects features based on a threshold for the smaller eigenvalue of the deformation and displacement matrix.

Lastly, the Lucas-Kanade optical flow algorithm, which has been cited some 20,000 times, was used for motion tracking (Lucas and Kanade, 1981). It is an image registration technique that utilizes spatial Gauss-Newton (Gauss, 1809) gradient descent non-linear optimization to direct the search for the feature that yields the best match. While a brute force approach would require quadratic complexity related to the dimensions of the feature, this method is linear in the number of pixels examined in the feature. An initial estimation of the disparity vector is updated via a Newton-Raphson iteration until convergence. The optical flow regression is done via computation of least squares (Legendre, 1805) with pixel weights based on a Gaussian function at the central pixel of the feature. The approach can handle feature rotation, scaling, and shearing.

Based on benchmarking with synthetic data, our algorithm had >99% detection accuracy and 98% tracking accuracy. The tracker formed <1% false positive links and about 2% of false negative links, however, these were due to a distance limit we introduced with the radius of the local search window. The development environment we used was of the cross-platform game engine of Unity Technologies (Unity, 2017). We compiled and tested our software on multiple platforms, such as Windows PC, Macintosh computer, iPhone smartphone, Android smartphone, and Microsoft HoloLens smart glasses. See Materials and Methods on how to download and test the software.

**Examples of real-time analysis**

A key novelty is that our computational platform outputs results in real-time, in parallel to image acquisition, which facilitates and speeds up research and clinical efforts by delivering instant visualization of data analysis and statistical outcomes. Such approach offers a clear improvement in the selection of imaging parameters and drug concentrations empirically by offering precise quantification of the changes in the dynamics of the underlying cellular processes. We present the functionality of our software with videos demonstrating specific features. Video 9 shows motion tracking of fluorescently labeled MT ends to measure polymerization rates in an epithelial cell. To compare subcellular differences in gene regulation, we have introduced the selection of a region of interest (ROI), shown as a red rectangle. Video 10 shows similar analysis, but with histograms for MT polymerization rates (the lower histogram) and for angular orientation of the MTs (the upper histogram). The graphics are overlaid on the original images and can be displayed as an augmented reality overlay in real time.

Video 11 shows glioblastoma multiforme tumor-propagating cells imaged with fluorescently labeled a histone H3K9me3 marker, which we tracked. Our analysis, which processed 19 frames per second, was focused on determining the nuclear reorganization in glial cells after drug treatment to reverse differentiation. Vectors with different shades of red show the angular orientation of the nuclear marks, similarly to the velocity vectors in Fig. 11. Vectors with different shades of blue show the speeds of motion, where lighter blue represents slower speeds.

Video 12 shows an example of our smartphone app, which detects MT ends and tracks MT motion when being pointed at a computer screen. The video shows the functionality during software development when the image dataset was used to calibrate the different modules of the software. Modules can be added and customized, depending on the application. Video 13 shows another example of our smartphone app, which detects MT ends and tracks MT motion when being pointed at a computer screen. In the video, molecular markers in cellular proteins encapsulated in a microfluidic droplet are being detected and tracked in real time. The tags appearing next to each EB1 comet in real time can either display the speed or the angle of motion, or indicate the identity number of each MT.

**Real-time motion tracking with transformers**

To extend the current functionality into obtaining the full trajectories, we will utilize the vectors obtained in each frame, together with the associated information on feature intensity and morphology, to generate embeddings (Vaswani et al., 2017) for a generative transformer network in which the tokens are the spatial coordinates of the features we track. This approach will allow training of the network to associate the most likely next feature in an image sequence, similarly to the way large language models (LLMs) generate text. Further, we will retrain a transformer network with a new set of tokens - with lists of 2D or 3D coordinates rather than words and with trajectories (lists of linked coordinates) rather than sentences of human speech, that is, we will retrain a LLM with motion trajectory data. We have generated several thousand biomedical time-lapse microscopy movies with about 2-10,000 trajectories consisting of about 3-20 protein features (either MT end binding proteins or fluorescent speckles (Ponti et al., 2003)) linked in each trajectory (Brennan et al., 2008; Galletti et al., 2014; Gatlin et al., 2010; Gatlin et al., 2009; Harkcom et al., 2014; Matov, 2024c; Matov, 2024d; Matov, 2024f; Matov et al., 2010; Matov et al., 2016; Matov et al., 2015; Thoma et al., 2010). These are about 60 million data points, but many more existing movies/data can be obtained from our academic collaborators. As we have pre-processed the data, the preliminary computer vision analysis has already been done and the results have been validated and published, which will facilitate the training and benchmarking.

**Disease classification with reinforcement learning**

One way to approach the classification of new patient samples is to utilize mathematical reasoning based on a large reasoning model (LRM) and reinforcement learning (RL) is utilized for post-training of LLMs and inference compute. Reasoning or a thought process consists of a sequence of tokens representing the steps in the reasoning trajectories. A Kullback-Leibrer (KL) divergence (Kullback and Leibler, 1951) constraint can achieve good empirical results on a range of challenging policy learning tasks (Schulman et al., 2015). The addition of a Group Relative Policy Optimization (GRPO) (Shao et al., 2024), a modification of the RL algorithm for Proximal Policy Optimization (Schulman et al., 2017), reduces the requirements for training data. For each new sample, GRPO samples a group of outputs from the old policy and then optimizes the policy model by maximizing the objective function, which is the expectation for the average reward of multiple sampled outputs, produced in response to the same question as the baseline. It utilizes a penalty based on computing the Monte Carlo approximations (Kungurtsev et al., 2023) of the KL divergence between the policy model and the reference model.

## DISCUSSION

RL techniques, such as Markov decision processes (Kakade and Langford, 2002; Schneider and Wagner, 1957), can provide the basis for sample stratification based on policies for sample classification in oncology as well as neurodegeneration. In particular, fluid intelligence (Cattell, 1943) applications of LRMs that depend on no prior training data can be useful in identifying and diagnosing new clinical conditions in real time. This context is important because many current disease types appear to be multiple diseases wrongly classified as one. The ability to distinguish different distinct clinical features will improve the precision of the treatment plans. Most misdiagnosed cases are the result of considering superficial clinical information in which indicators for different conditions happen to overlap. Without distinct markers, the diagnosis is made based on the highest probability, given incomplete information and the historical data – which systematically undermines the statistical representation of some diseases to the point of considering them as rare (Marwaha et al., 2022) diseases.

Further, our approach will facilitate the creation of a medical digital twin (Björnsson et al., 2019; Eddy and Schlessinger, 2003; Laubenbacher et al., 2024; National Academies of Sciences et al., 2023; Sadée et al., 2025). A digital twin can be broken down into the following five essential components – physical object (patient), data connection, model (patient-in-silico), interface, and twin synchronization. A medical digital twin focuses on the patient's health, as described by multimodal data – including, but not limited to, laboratory tests, imaging, sequencing, and the electronic health records (EHRs).

In cases with localized medical issues, the patient becomes the specific organ until a broader context is required. Patient-derived organoids can serve as organ and disease models. Novel technologies that merge AI approaches and mechanistic models offer enormous promise in generating a high-fidelity patient-in-silico model. The interface allows the clinical team to establish the effect of different therapies, query the patient-in-silico, and subsequently integrate its findings into the EHR. The findings are evaluated by uncertainty quantification (Chan and Elsheikh, 2018; Elsheikh et al., 2012; Sumner et al., 2012), which tracks the accumulation of different errors from data acquisition to modelling. Synchronization, based on liquid biopsy and body fluids testing, takes place when recording a substantial deviation from the baseline or measuring the effects of patient treatment. For treatments of diseases such as cancer, twin synchronization might involve updating the medical digital twin after each chemotherapy cycle.

AI can generate predictive models even if the underlying disease mechanisms are unknown. For instance, recurrent neural networks (RNNs) have been used to predict the accumulation of somatic mutations in cancer cells, with implications for prognosis and treatment strategy (Auslander et al., 2019). In another example, RNNs were used to predict changes in lung tumor shape throughout a radiation treatment cycle and mitigate exposure to healthy tissue (Li et al., 2021). Mechanistic models are used to model the changes in insulin receptors in diabetes, tumor growth in cancer, or the function of entire organs (Baillargeon et al., 2014; Metzcar et al., 2019; Rathee and Nilam, 2017; Tikenoğulları et al., 2022), but simplifications and assumptions lead to inaccuracies. The quantitative live-cell image analysis approach we propose can elucidate the underlying disease mechanisms based on the interrogation of patient-derived cells and provide important novel information to be integrated in the AI and mechanistic models.

By generating a highly detailed and personalized disease model, the patient-in-silico that can be used to generate patient disease metrics, predict disease progression, and simulate different treatment outcomes. Image archiving and communication systems (Choplin et al., 1992) facilitate the aggregation of datasets from different imaging modalities as well as data fusion and harmonization (Nan et al., 2022). To update the disease models, new patient cells for live-cell cultures and sequencing can be obtained from body fluids (Matov, 2024a; Matov, 2024j) and liquid biopsies (Sung et al., 2013; Sung et al., 2012) after each round of therapy. This way, with the addition of quantitative live-cell image analysis datasets, our approach to diagnostics will contribute to translating the promising new paradigm of a medical digital twin from theory into clinical practice.

Quantitative cell biological profiling of subcellular interactions in disease is required to understand the mechanisms of pathogenesis and elucidate the mechanisms of drug action. During pathogenesis, or due to drug treatment in disease, cells change their intracellular organization, rearrange their internal components as they grow, divide, and adapt mechanically to a hostile environment. These functions depend on proteins, cytoskeletal filaments, and FA complexes, which provide the cell shape and its capacity for directed movement. The integration of measurements of intracellular dynamics and the contribution of multiple genetic pathways in degenerative diseases is vital for the development of biomarkers for the early detection of pathogenesis and therapy efficacy. Methods that can reliably analyze the evolution of the morphology and localization of cellular proteins over hundreds of time-lapse frames will be very relevant for capturing the changes in the subcellular organization in disease and during treatment. They will allow physicians to compare visually and quantitatively the effects of treatment regimens and select the one most likely to be efficacious.

The RAS family of GTPases (KRAS, NRAS, HRAS) regulates cell proliferation and survival by cycling between active, GTP-bound (ON) and inactive, GDP-bound (OFF) states (Simanshu et al., 2017). The KRAS$^{G12C}$ selective inhibitors sotorasib and adagrasib cannot target GTP-bound RAS and that limits their clinical utility (Kim et al., 2023; Kim et al., 2020). TCIs represent a novel class of therapeutics developed in order to overcome the limitations by targeting GTP-bound RAS (Araujo et al., 2024; Schulze et al., 2023; Wasko et al., 2024) and restoring GTP hydrolysis (Cuevas-Navarro et al., 2025). TCIs recruit the intracellular protein cyclophilin A (CYPA) to GTP-bound RAS. Elironrasib (RMC-6291) and zoldonrasib (RMC-9805) covalently target KRAS G12C and G12D, respectively; daraxonrasib (RMC-6236) targets multiple RAS variants in a reversible manner (Holderfield et al., 2024; Jiang et al., 2023; Schulze et al., 2023). Acquired drug resistance to TCIs involves mechanisms of preventing drug-target engagement and inducing RAF dimerization (Sang et al., 2025). One of the resistance mechanisms is acquired MEK1 mutations (Sang et al., 2025). The activity of MEK1 destabilizes interphase MTs and reduces mitotic spindle defects (Tolg et al., 2010). Furthermore, CYPA binds cytoplasmic dynein and co-localizes with MTs (Galigniana et al., 2004), suggesting the utilization of combination therapy that includes drugs modulating the activity (Garg and Alisaraie, 2025) of the MT minus end-directed motor dynein (Höing et al., 2018) to overcome drug resistance. Quantitative live-cell microscopy analysis of MT dynamics can facilitate the optimization of solid tumor therapy.

Tumor growth curves can be quantified using mathematical models, such as the Gometz function (Gompertz, 1825; Winsor, 1932). Heuristic analysis (Jungius, 1622) brought sigmoid growth kinetics into the arena of cancer therapeutics by establishing the observation that a given administration of cancer therapy killed a specific fraction of the cancer cells present rather than an absolute number (Skipper et al., 1964). This observation inherently leads to the conclusion that acquired drug resistance is inevitable. Emerging evidence implicates nonmutational mechanisms, including changes in cell state during the early stages of acquired drug resistance. Targeting nonmutational resistance may, therefore, present a therapeutic opportunity to eliminate residual surviving tumor cells and impede relapse (Hangauer et al., 2017).

The integration of measurements of intracellular dynamics and the contribution of multiple genetic pathways in degenerative diseases is vital for the development of biomarkers for the early detection of pathogenesis and therapy efficacy. We developed a software suite (DataSet Tracker) for real-time analysis designed to run on PCs, Macs, smartphones, and smart glasses and suitable for resource-constrained (Khosrow-Pour, 2005), on-the-fly (Happe et al., 2013) computing in microscopes (Edelstein et al., 2010) without internet connectivity; a demo is available for viewing in Vid. 1. Our objective has been to present the community with an integrated, easy to use by all, tool for resolving the complex cytoskeletal dynamics and it is our goal to have such software system approved for use in the clinical practice. A key novelty is that our computational platform outputs results in real time, during imaging and without having to store the data first, which will facilitate and significantly speed up research and clinical efforts by providing instant delivery of data analysis. Augmented reality AI software (Chen et al., 2019) can be added to existing live-cell microscopes, thus providing instant visual feedback with added graphics during sample observation. It will enhance the images by also displaying numbers describing the measured differences and, in doing so, facilitate the image interpretation by the human operator. This functionality will significantly enhance the ability of clinicians to make quick and correct decisions regarding drug action.

There exist very few disease biomarkers allowing for longitudinal *in vivo* analysis like fluorescence (Delori et al., 1995) puncta in the fundus of the eye. As an extension of the central nervous system (De Groef and Cordeiro, 2018), the eye harbors retina ganglion cells vulnerable to degeneration, and visual symptoms are often an early manifestation of neurodegeneration (Vujosevic et al., 2023). Imaging the retina, following an intravenous injection of an ocular tracer (Aguilar-Calvo et al., 2022; Pilotte et al., 2024; Pilotte et al., 2025), allows performing minimally invasive *in vivo* imaging in patients with a number (over a dozen) of neurodegenerative diseases originating in different areas of the central nervous system. Time-lapse imaging during drug treatment allows for the comparison of the effects of drug regimens on their target proteins. Such companion diagnostics (Agarwal et al., 2015) allows to evaluate the relative contribution of different misfolded or aggregated proteins in drug resistance and response; it also allows the discovery of predictive and prognostic biomarkers for the early detection of neurodegeneration and advanced patient stratification analysis. During *in vivo* analysis, retina imaging is associated with rapid three-dimensional rotation but with on-the-fly registration (Hong et al., 2024), a quantitative readout of phenotype changes associated with disease can be delineated in real time. Action recognition algorithms can deliver image segmentation (Hou et al., 2019; Sun et al., 2024) and feature metrics as the basis for predictive and prognostic biomarkers.

Real-time image analysis can also assist surgeons with lesion localization during surgical resections (Deivasigamani et al., 2023). The software will segment the precise area of selected fiducial biomarkers on-the-fly, enhance the image features of interest, and analyze the time evolution of their contour (Matov, 2024h) to improve the precision of the resection boundary. Similarly, it will improve the laser positioning guidance during multi-parametric MRI-guided focal laser ablation therapy (van Luijtelaar et al., 2019). We can envisage many additional applications of our software system, for instance in ultrasound imaging (Rezai et al., 2024). On-the-fly image contrast enhancement and feature contour display (Buda et al., 2023), together with the addition of automated aerial metrics on the screen, will improve the precision of the measurements and the diagnosis.

The complexity of current drug regimens in oncology highlights the incredible ability of the neoplastic cell to survive. It suggests the need to develop (easy to use in the clinic) quantitative methods to evaluate drug treatments' short-term and long-term effects on their cellular target. Often when tumors respond to treatment, the disease eventually returns and, in most cases, does respond to other therapies, even though resistant to the previously given one. This type of drug response indicates that it is crucial to develop tools for optimal drug selection upfront and avoid intrinsic drug resistance as well as anticipate the effects of tumor plasticity and the development of acquired drug resistance (Groenendijk and Bernards, 2014; Hangauer et al., 2017). Heterogeneous tumors (Network, 2015) evolve as a result of drug treatment. During hormonal therapy, cells acquire a loss of the surface expression of their receptor ligands (Formaggio et al., 2021; Zattarin et al., 2020). Very often the drug treatment process would effectively serve as a repeated selection step, allowing for the replication of the most resilient clones. The surviving clone often lacks any known molecular biomarkers for targeted therapy, thus leaving systemic therapy, associated with high toxicity and low cure rates, as, perhaps, the only option (Galletti et al., 2014; Petrylak et al., 2004; Tannock et al., 2004). However, drug regimens approved for other indications can, sometimes serendipitously (Linn et al., 1994), induce a full remission and libraries with FDA-approved compounds can routinely be tested in *ex vivo* or *in vivo* quantitative imaging assays to evaluate their drug-target engagement (Thadani-Mulero et al., 2014) and compare their off-label efficacy. Our ability to observe such events and "funnel (Franck, 1541; Harsdörffer, 1647) something in" is very limited compared to the rates with which we acquire new information (Fig. 12, taken from (Fontolliet, 2001)) by looking at imaging datasets. Therefore, our computational approach can lead to the discovery of putative, secondary mechanisms, which have thus far been not appreciated.

Upon an extensive clinical validation, besides anticipating drug resistance, quantitative imaging systems can be utilized to identify sensitizing drugs – targeting the cytoskeleton – to treat disease with known resistant genetic profiles. Further, it will be beneficial to analyze the changes of the cellular phenotype of patient-derived cells of ultimate responders. This way, it will become feasible for quantitative microscopy methods to be used in selecting regimens that achieve complete response and elimination of residual disease.

Given the availability of high-quality fundamental research equipment for high resolution live-cell microscopy in most university hospitals, we propose to embed in these systems real-time augmented reality software for automated quantification and on-the-fly statistical analysis of intracellular behavior. Introducing the quantitative imaging method to the clinic will allow physicians to fine-tune, with a high level of certainty, an optimal treatment regimen for each patient.

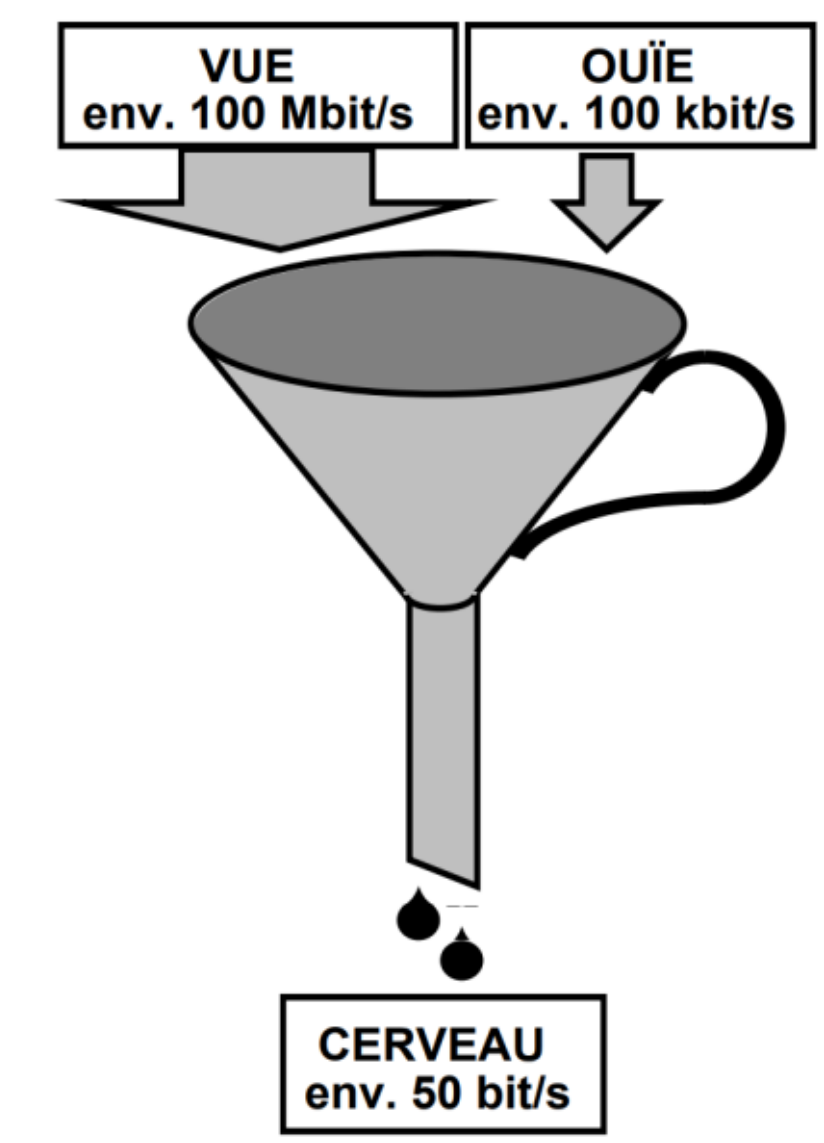

**Figure 12. The real Nuremberg funnel.** The speed of information flow through our eyes (about 100 Mbit/s) and ears (about 100 kbit/s) is far greater than the speed with which our brain can process it (about 50 bit/s).

The main concern regarding patient treatment, which motivated this work, has been that important differential effects of therapies on their cellular targets pertaining to drug efficacy cannot reliably be identified with genetic sequencing and pathology slides alone. There are differences in the mechanisms with which cells react to treatment, for example in regard to the immune system, which can only be appreciated by live-cell imaging methods (Mitchison, 2021). Further, the effects of a physiologically relevant dose (Matov, 2024f) of a drug are not visible to the naked eye by looking through the microscope lens, in part, because of the inability of our vision to detect a gradient in pixel intensity when bright features are co-localized with dim features (Aubert, 1865). In these and many other instances, the cell biology imaging methods require quantification because of the overwhelming number of cellular proteins and components involved. For these reasons, we consider real-time image analysis software as an indispensable tool to extract a truthful readout of the underlying pathological processes.

The potential clinical application of the analytics approach outlined in this manuscript pertains to anticipating drug resistance in cancer therapy (Vasan et al., 2019) and the treatment of neurodegeneration (Ahmad et al., 2021) based on the microscopic evaluation of living patient cells *ex vivo*. Until now, the accurate computational analysis of dynamic cytoskeletal structures was limited due to the lack of appropriate software tools. Since the entangled protein networks in the cell are fast evolving, rapidly changing their turnover rates and directions of motion, it has been a very tedious process for scientists using traditional motion tracking methods to resolve complex motion patterns (Matov, 2024h). Our method, by not requiring any additional steps for the analysis - but rather displaying the tracking vectors in real time - can aid in establishing a new level of insight into cellular processes and, in doing so, advance our understanding of the dynamic organization of the cytoskeleton, cell division and motility. In this context, it is our goal to advance the field of drug development and contribute to improving patient care.

## MATERIALS AND METHODS

### Image analysis

All image analysis programs for detection and tracking of comets, and graphical representation of the results were developed in Matlab and C/C++. The EB1 comet analysis method ClusterTrack used is described and validated in (Matov et al., 2010). The computer code is available for download at: https://www.github.com/amatov/ClusterTrackTubuline.

The image analysis programs for real-time detection and tracking, and graphical representation of the results were developed in the cross-platform game engine Unity in C#. OpenCV for Unity requires a license from Enox Software. The computer code is available for download at: https://www.github.com/amatov/DataSetTracker.

To test the software, we have uploaded EBTracker.exe and DataSetTracker_v1.1.zip files at the same link. In the zip folder, there are instructions on how to test the tracker. It does not require any installations and starts on double click. A default test video is included with the download and the analysis begins upon clicking the software executable file. The user can change the video for analysis.

IFTA, which further describes spot detection algorithms, is described in (Matov, 2024b). The computer code is available for download at: https://www.github.com/amatov/InstantaneousFlowTracker.

The 10x image analysis programs for PSMA, α-tubulin, DAPI, and CD45 segmentation as well as 3D AR and DAPI segmentation, and graphical representation of the results were developed in Matlab and C/C++. The wavelet transform method used, spotDetector, was described and validated in (Olivo-Marin, 2002) and the unimodal pixel intensity thresholding in (Rosin, 2001). We identified the CTC areas as connected-component labeling pixel lists in the epithelial tumor imaging channel (such as PSMA or occasionally cytokeratin (CK)) for which in the nuclear imaging channel there is a DAPI stain with a statistically representative size and a circular shape. At the same time, we required that these cells are negative in the CD45, that is, the absence of a leukocyte marker. Similarly, to detect neutrophils and lymphocytes, we identified clusters of bright pixels in the CD45 channel for which a nuclear area is detected in the DAPI channel and the epithelial stain is not present. On multiple occasions, we detected double-positive (PSMA+/DAPI+/CD45+ or CK+/DAPI+/CD45+) and double-negative (PSMA-/DAPI+ /CD45- or CK-/DAPI+/CD45-) cells, which we classified in separate bins.

Drug-induced MT bundles were evaluated in terms of thickness and texture parameters in 63x images. The detection of tyrosinated tubulin in 10x images was considered to be a marker of drug sensitivity. We analyzed the cellular localization of AR in 3D image stacks, in which we performed segmentation in 35 individual 2D images and reconstructed the volume, and classified the CTCs as sensitive to drug treatment if the AR is within the volume of the nucleus and resistant if the AR is mainly in the cytoplasm.

Cell membrane areas in with TGFβ-containing secretory granules on the cell surface were identified by image foreground segmentation based on Gaussian filtering (Weierstrass, 1885) with a high value of the standard deviation in order to retain the areas with the fluorescent granules. A unimodal pixel intensity thresholding (Rosin, 2001) was then applied and pixels below the threshold were computationally clipped and set to zero. The resulting images are filtered with a Gaussian kernel with frequency cut-off matched with the limit of the optical transfer function (Fourier, 1822) beyond which the collected signal is white noise and the kernel was computationally fitted to the Bessel function (Bernoulli, 1728) forming the point-spread function of the imaging system. The granules were counted as the number of detected spots as clusters of bright pixels within the area of the segmented cell outline. To this end, we applied a method introduced in (Olivo-Marin, 2002) to detect bright spots in fluorescence images based on the multiscale product of subband images resulting from the à trous wavelet transform (Starck et al., 2000) decomposition of the original image, after thresholding of non-significant coefficients. The multiscale correlation of the altered wavelet coefficients, which allows enhancing multiscale peaks due to spots, while reducing noise, combines information coming from different levels of resolution and gives a clear and distinctive characterization of the spots.

To calculate the overall intensity in immunofluorescence images of MTs in cells growing in clusters, we identify the centroid position of the nucleus by applying unimodal pixel intensity thresholding (Rosin, 2001) in the DAPI channel, which allows the segmentation of the nuclei. We then use the centroids to build a Voronoi diagram (Dirichlet, 1850) in the MT channel, which provides the watershed lines dividing the individual cells. The computer code is available for download at:
https://github.com/amatov/SegmentationBiomarkerCTC,
https://github.com/amatov/ResistanceBiomarkerAnalysis,
https://github.com/amatov/AntibodyTextureMorphology.

To match baseline AF puncta in fundus images with hyper-fluorescent puncta after injection with fluorescence emitting dye, we utilize SURF (Bay et al., 2008). In brief, SURF is a scale space-based (Lindeberg, 1993) transform applied at multiple scale for feature detection. For scale selection, the determinant of the Hessian matrix (Hesse, 1872) of second-order partial image derivatives is computed. Features are selected at different scales and Gaussian filtering (Weierstrass, 1885). Feature matching is based on a distance between the descriptor vectors, for example, the Mahalanobis (Mahalanobis, 1936) or Euclidean distance. The SURF descriptor is computed for a circular region around each feature and consists of intensity distribution of the pixels and feature orientation based on Haar wavelet transform (Haar, 1910) or Gabor filter (Gabor, 1946). The code for analysis of fundus fluorescence is available at:
https://github.com/amatov/NeurodegenerationFluorescentFundus.

**Bioinformatics analysis**

Sequence reads were trimmed using Trimmomatic 0.36 (Bolger et al., 2014). We used hg38 genome as reference and aligned the reads with BWA-MEM version 0.7.15 (Li and Durbin, 2010). For variant calling, we utilized SAMtool (Li et al., 2009), BCFtools (Li, 2011), Ensembl Variant Effect Predictor v86 (McLaren et al., 2016), VCFtools (Danecek et al., 2011), and ANNOVAR (Wang et al., 2010). To run the analysis, we installed a virtual machine (Oracle VM Virtual Box 5.0.6) and Ubuntu 12.04.5 on a Windows server with Intel Xeon E5-1620 processor with 4 dual cores and a solid-state drive. For the alignment step, we connected to a high performance computing cluster (Mount Moran, 2016).

**Sample processing**

Retroperitoneal lymph node organoid whole genome sequencing (736 ng DNA) was done at the UCSF Core Facilities.

**Cell culture**

Gastric cancer cell lines SCH, MKN7, HS746T, FU97, AZ521, and TMK1 were a gift from Patrick Tan; MKN1 and SNU1 were a gift from Zev Wainberg. All other cell lines were obtained from ATCC.

All breast cancer cell lines were cultured in DMEM (CellGro, Mediatech) supplemented with 10% FBS (Atlanta Biologicals), incubated at 37°C in 5% $CO_2$. MT polymerization dynamics was visualized by lentivirus-mediated low level expression of EB1ΔC-2xEGFP (Matov, 2024f), a marker of growing MT plus ends that does not interfere with endogenous EB1 binding partners. The MDA-MB-231 TNBC cells were grown also as 3D cultures in Matrigel (Corning) with MT polymerization dynamics visualized by lentivirus-mediated low level expression of EB1ΔC-2xERFP.

Primary and retroperitoneal lymph node metastatic prostate tumor and sternum metastatic rectal tumor tissues were dissociated to single cells using modified protocols from the Witte lab (Goldstein et al., 2011). Organoids were seeded as single cells in three 30 µL Matrigel drops in 6-well plates. Organoid medium was prepared according to modified protocols from the Clevers lab (van de Wetering et al., 2015) and the Chen lab (Gao et al., 2014).

To obtain a single cell suspension, tissues were mechanically disrupted and digested with 5 mg/mL collagenase in advanced DMEM/F12 tissue culture medium for several hours (between 2 and 12 hours, depending on the biopsy and resection performed). If this step yielded too much contamination with non-epithelial cells, for instance during processing of primary prostate tumors, the protocol incorporated additional washes and red blood cell lysis (Goldstein et al., 2011). Single cells were then counted using a hemocytometer to estimate the number of tumor cells in the sample, and seeded in growth factor-reduced Matrigel drops overlaid with prostate cancer medium (Gao et al., 2014). With radical prostatectomy specimens, we had good success with seeding 3,000 cells per 30 µL Matrigel drop, but for metastatic samples organoid seeding could reliably be accomplished with significantly fewer cells, in the hundreds. To derive organoids from patient CTCs, liquid biopsy samples of 40 mL peripheral blood would be collected, processed, and plated in a Matrigel-Collagen-Fibronectin matrix to form organoids similarly to the metastatic breast cancer organoids we cultured from mouse CTCs (Matov, 2024c).

To transduce organoids, we modified protocols from the Clevers lab (Koo et al., 2013) to adapt to the specifics of prostate organoid culture (such as the significant differences in proliferation rates in comparison to colon and rectal organoids). We found out empirically that cells in mid-size organoids (60-100 µm in diameter) infect at much better rates than trypsinized single organoid cells, small organoids, or large organoids. These were the steps we followed to express EB1ΔC-2xEGFP in organoids: (1) Add Dispase (1 mg/mL) to each well to dissolve the Matrigel at room temperature for 1 hour. (2) Spin down (at 1,000 rpm for 4 minutes) and mix organoids with 10 µL of viral particles (enough for 1 well with three Matrigel drops of 30-40 µL with organoids containing 1-2 million cells) with 10 mg/mL Y27632 ROCK inhibitor and 10 mg/mL polybrene for 30 minutes. (3) Spin the organoids with viral particles for 1 hour at 600 g. (4) Leave the organoids for a 6-hour incubation. (5) Spin down and plate in Matrigel. (6) 1 hour later, add medium. We used 0.5 mg/mL blasticidin for only one round of medium (3 days) because an increase of the density of labeled cells in the organoids reduced our ability to image MT tips with good contrast.

MDA-MB-231 and SK-BR-3 cells were treated with paclitaxel and docetaxel for 2 hours in titration experiments (0.023 nM, 0.69 nM, 0.206 nM, 0.617 nM, 1.852 nM, 5.556 nM, 16.667 nM, 50 nM, and 150 nM) to perform a sulforhodamine B cytotoxicity assay.

MDA-MB-231 cells were pre-treated with 0.1% DMSO (vehicle control) or 100 nM paclitaxel for 6 hours, washed in serum-free DMEM, then treated with 100 µg/mL LPS in serum-free DMEM for 2 hours to induce TGFβ trafficking to the cell membrane. Cells were fixed in PHEMO buffer (Carbonaro et al., 2011) without any detergent to ensure that cells remained non-permeabilized, then immunostained with a rabbit anti-TGFβ antibody, followed by Alexa Fluor 488-conjugated secondary. The total cell area was visualized by counter-staining with a cell membrane dye. TGFβ foci were quantified using an adaptation of our blob detection algorithm (Matov, 2024b; Matov et al., 2011). LPS derived from *Escherichia coli* O111:B4 was obtained from Sigma-Aldrich. Paclitaxel (in intravenous solution) was a gift from Linda Vahdat. Docetaxel was obtained from Sigma-Aldrich in powder form.

**Microscopy imaging**

MDA-MB-231 cells were transfected with plasmid encoding EB1ΔC-2xEGFP (Piehl and Cassimeris, 2003) and EB1 comets in live cells were imaged by spinning-disc confocal microscopy using a 100x magnification oil immersion 1.49 NA objective for all cultured cells as previously described (Gierke and Wittmann, 2012) and tracked using our image analysis algorithm (Matov et al., 2010). EB1 transiently binds to growing MT plus ends (Akhmanova and Steinmetz, 2008), generating a punctate pattern of EB1ΔC-2xEGFP comets throughout the cell. The exponential decay of available binding sites results in the characteristic comet-like fluorescence profiles of EGFP-tagged end binding proteins ("comets").

Organoids were imaged using transmitted light microscopy at 4x magnification and phase contrast microscopy at 20x magnification on a Nikon Eclipse Ti system with camera Photometrics CoolSnap HQ2.

**Abbreviations**

AF – auto-fluorescence

AI – artificial intelligence

AR – androgen receptor

BRAF – serine/threonine-protein kinase B-Raf

BRCA1/2 – breast cancer type 1/2 susceptibility protein

B1KD – A549 lung cancer cells after knockdown of BRCA1

CD45 – protein tyrosine phosphatase, receptor type, C

CK – cytokeratin

CLIP170 – cytoplasmic linker protein, 170

CTGF – connective tissue growth factor

COX – prostaglandin-endoperoxide synthase

CTC – circulating tumor cell

CYPA – cyclophilin A

c-Met – hepatocyte growth factor receptor

DMSO – dimethyl sulfoxide

DAPI – 4′,6-diamidino-2-phenylindole

EB1/3 – microtubule end binding protein, 1/3

EB1ΔC – EB1 construct truncated at amino acid 248 that does not interact with other proteins at the microtubule end

EGFP – enhanced green fluorescent protein (2xEGFP indicates two molecules)

EHR – electronic health record

FA – focal adhesion

F-actin – filamentous actin

GRPO – group relative policy optimization

GSK3β – glycogen synthase kinase 3, beta isoform

GTP – guanosine triphosphate

IFTA – instantaneous flow tracking algorithm

MT – microtubule

NSCLC – non-small cell lung cancer

LPS – lipopolysaccharide

LLM – large language model

LRM – large reasoning model

MRI – magnetic resonance imaging

NuMA – nuclear mitotic apparatus protein

OpenCV – open source computer vision library

PC – personal computer

PSMA – prostate-specific membrane antigen (muJ591 – anti-PSMA antibody)

RL – reinforcement learning

RNN – recurrent neural network

ROCK – Rho-associated coiled-coil kinase

ROI – region of interest

SURF – speeded-up robust features

TBK1 – TANK binding kinase, 1

TCI – tri-complex inhibitor

TGFβ – transforming growth factor, beta isoform

**Ethics declaration**

IRB (IRCM-2019-201, IRB DS-NA-001) of the Institute of Regenerative and Cellular Medicine.

Ethical approval was given.

Approval of tissue requests #14-04 and #16-05 to the UCSF Cancer Center Tissue Core and the Genitourinary Oncology Program was given.

Informed consent to participate was obtained for clinical studies with IRB protocols 0804009740 and 0707009283.

Informed consent was obtained from all subjects and/or their legal guardian(s). The study adhered to the Declaration of Helsinki.

## Consent for publication

Consent for publication was given.

## Author contributions

A.M. conceived the study, performed cell culture, microscopy imaging, and data analysis, prepared the figures, and wrote the manuscript.

## Data availability statement

The datasets used and/or analyzed during the current study are available from the corresponding author upon reasonable request. The prostate cancer organoid whole genome sequencing datasets (https://www.ncbi.nlm.nih.gov/bioproject/1255230) are accessible at: https://www.ncbi.nlm.nih.gov/sra/SRX28551163.

## Conflicts of interest

The author declares no conflicts of interest.

## Funding statement

No funding was received to assist with the preparation of this manuscript.

## ACKNOWLEDGEMENTS

I thank James Cumberbatch (Vidro Ltd) for his help with Unity and XCode, and Neil Bander for the PSMA data. Giuseppe Galletti performed the $IC_{50}$ experiments. The patient blood samples analyzed were from clinical studies with IRB protocols 0804009740 and 0707009283 at Cornell Medicine. I am grateful to the Genitourinary Tissue Utilization committee and the Genitourinary and Prostate SPORE Tissue Cores at the UCSF Cancer Center for the approval of my tissue requests #14-04 and #16-05, the Stand Up To Cancer / Prostate Cancer Foundation (SU2C/PCF) West Coast Dream Team (WCDT), the Institute of Regenerative and Cellular Medicine for issuing the Institutional Review Board protocol approval IRCM-2019-201, IRB DS-NA-001 for the observational study "Longitudinal analysis of next-generation sequencing of nucleic acids for early detection of degenerative diseases such as cardiovascular, neoplastic and diseases related to the nervous system", and James Faber for his feedback regarding the protocol and the process of approval. A preprint of this paper is available at arXiv (Matov, 2024i).

## SUPPLEMENTARY MATERIALS

Video 1 – A demo video of the software functionality in high resolution.

https://vimeo.com/manage/videos/238144544

Video 2 – Prostate cancer M12 cell expressing AR-wild type and MT labeling with overlaid EB1 comet detections. https://vimeo.com/1060977612/ce83b5bd52

Video 3 – Prostate cancer M12 cell expressing AR-V7 and MT labeling with overlaid EB1 comet detections. https://vimeo.com/1060978105/1cd7d9779b

Video 4 – Castrate-resistant prostate cancer organoid cell with MT labeling and EB1 comets. Organoids derived from acetabulum metastasis. https://vimeo.com/1060978355/812e82c658

Video 5 – Multiple myeloma exosomes in MM1S cells with labeled CD63.
https://vimeo.com/1060982925/1c6381fb1f

Video 6 – Synthetic movie with moving bright features. https://vimeo.com/999588708/43e6111879

Video 7 – Overlay of Video 6 with vectors displaying the tracking results.
https://vimeo.com/999589508/d6dbcf35f5

Video 8 – Detection of synthetic markers with frame by frame tracks.
https://vimeo.com/1058382285/00f06c9bb1

Video 9 – Tracking MT ends in epithelial cells with a ROI selection. The red rectangle shows the selected ROI. https://vimeo.com/1058385496/8299567e00

Video 10 – Tracking MT ends in epithelial cells with augmented reality graphics. Histograms showing the distributions of protein speeds and angular orientation are added on the screen during processing. https://vimeo.com/1058418895/5c6ecf2821

Video 11 – Tracking H3K9me3 in glia cells for glioblastoma drug discovery. Epigenetic histone markers were labeled and track to investigate nuclear reorganization after drug treatment. Shades of red denote the direction of the vectors. Shades of blue denote the different speeds. 19 frames per second are processed. https://vimeo.com/1058419537/07b1920b97

Video 12 – Phone app real-time MT ends tracking calibration. The video shows testing the phone app during software development. https://vimeo.com/1058389251/7d6603a5e1

Video 13 – Tracking MT ends via a phone app in real time. Detection and tracking is done in real time by pointing the phone toward the computer screen. https://vimeo.com/1058387386/cdbe5c3d74

A software folder 'DataSetTracker_v1.1' contains the DataSet Tracker executable file, a test video, and a readme file with instructions. You may also download DataSetTracker_v1.1.zip and EBTracker.exe from https://github.com/amatov/DataSetTracker. In brief, place EBTracker.exe in the unzipped 'DataSetTracker_v1.1' folder and upon double click it will start the real-time analysis of the default MP4 video located in folder 'Streaming assets' in 'EBTracker_Data'.